\documentclass[aps,prd,floatfix,superscriptaddress,twocolumn,showpacs,amssymb]{revtex4-1}
\usepackage{amssymb,amsmath,verbatim}
\usepackage{graphicx}
\usepackage{epsfig}
\usepackage{dcolumn}
\usepackage{bm}

\newcommand{\be}{\begin{equation}}
\newcommand{\ee}{\end{equation}\noindent}
\newcommand{\eei}{\end{equation}}
\newcommand{\bea}{\begin{eqnarray}}
\newcommand{\eea}{\end{eqnarray}\noindent}
\newcommand{\eeai}{\end{eqnarray}}

\newcommand{\hf} {\frac{1}{2}}
\newcommand{\nn}{\nonumber\\}

\def\eq#1{(\ref{#1})}
\def\la{\langle}
\def\ra{\rangle}
\def\ord#1{{\cal O}(#1)}

\def\t#1{{\tilde#1}}
\def\c#1{{\cal#1}}
\def\b#1{{\bar#1}}

\def\tr{{\mathrm{tr}}}

\def\u#1{{\underline{#1}}}

\def\fdd#1#2#3{\frac{\delta^2#1}{\delta#2\delta#3}}

\begin{document}
\title{Phase structure of the  $O(2)$ ghost model with higher-order gradient term
}

\author{Z. P\'eli}
\affiliation{Department of Theoretical Physics, University of Debrecen,
P.O. Box 5, H-4010 Debrecen, Hungary}

\author{S. Nagy}
\affiliation{Department of Theoretical Physics, University of Debrecen,
P.O. Box 5, H-4010 Debrecen, Hungary}
%\affiliation{MTA-DE Particle Physics Research Group, P.O.Box 51, H-4001 Debrecen, Hungary}
\author{K. Sailer}
\affiliation{Department of Theoretical Physics, University of Debrecen,
P.O. Box 5, H-4010 Debrecen, Hungary}

\date{\today}

\begin{abstract}
The phase structure and the infrared behaviour of the Euclidean 3-dimensional $O(2)$ symmetric ghost scalar field model with higher-order derivative term has been investigated in Wegner and Houghton's renormalization group framework. The symmetric phase in which no ghost condensation occurs and the phase with restored symmetry but with a transient presence of a ghost condensate have been identified. Finiteness of the correlation length at the phase boundary hints  to a  phase transition  of first order. The results are compared with those for the ordinary $O(2)$ symmetric scalar field model. 

\end{abstract}



\maketitle

\section{Introduction}

Scalar fields with negative quadratic gradient term may have relevance in
various gravity theories as the dilaton field in Einstein, Einstein-Hilbert and Einstein-Maxwell gravity as well as  gravitating phantom scalar fields coupled to gravity and/or the electromagnetic field. The dilaton field, being a one-component real scalar field, occurs by making the conformal degree of freedom of the metric explicit. The usual kinetic term of the dilaton field as well as that of the phantom fields is of `wrong' sign and this implies that those fields are repulsively coupled to gravity. In quantum field theory  the phantom fields are the ghost fields.

A great amount of astronomical observational data collected from type Ia supernovae, large scale structures, and cosmic microwave background anisotropy support
that our Universe is under accelerated expansion \cite{Rie1998,Teg2004,Spe2003}. There is observational evidence \cite{Ton2003} that at about 70 percent of the contents of our Universe consists of dark energy, some exotic energy with negative pressure that is reponsible for the accelerated expansion by its gravitational repulsion. In various cosmological models the  dilaton field and the gravitating 
 phantom scalar fields are possible candidates for producing dark energy \cite{Ami2015,Bam2013}. Also the cosmological evolution in scalar-tensor gravity may
 show up phantom epochs that occur as a result of dynamics \cite{Bar2007}.
 Nonstationary models for gravitational collapse of gravitating phantom fields and formation of phantom black holes have been widely investigated recently \cite{Nak2013}, including the description of the internal structure of phantom black holes \cite{Azr2013}, their thermodynamical properties \cite{Rod2012}, as well as the consequences of the no-hair theorem stating that perturbations outside of the black hole are either gravitationally radiated away to infinity or swallowed by the black hole \cite{Gra2014}. 

 It is the main goal of the present paper to determine the phase structure
and the infrared scaling behaviour
of the Euclidean 3-dimensional $\phi^4$ scalar model with internal $O(2)$  symmetry in the particular case when it possesses the kinetic energy operator $\Omega(-\Box)=-Z\Box
+ Y \Box^2$ with negative wavefunction renormalization constant, $Z<0$ ($\Box=\partial_\mu\partial_\mu$ denotes the  d'Alembert operator).

Scalar fields with the `wrong' sign of the usual kinetic term (like the dilaton field and phantom scalar fields) may condensate when suitable higher-order gradient terms  are present in the model \cite{Arka2004}. The situation is similar to the case of
the spontaneously symmetry broken phase of usual $\phi^4$ theory, when the scalar field acquires a nonvanishing vacuum expectation value and the contribution of
the small fluctuations around the new minimum of the energy remains bounded. Gradient terms in the  Lagrangean of the type $\phi \Omega(-\Box) \phi$ with $Z<0$
and $Y>0$ may provide an inhomogeneous condensate with deeper minimum of energy as that would be for a homogeneous field configuration.
Even for such theories in Euclidean space, the stationary-wave modes of the field $\phi$ with momenta $p^2\le -Z/Y$ have  negative kinetic contribution to the action, and the corresponding spatially inhomogeneous configurations may produce
a deeper minimum of the action than that would be produced by any homogeneous field configuration minimizing the potential. For a detailed analysis of the ghost-condensation mechanism see e.g.  \cite{Lau2000,Bon2013}.
 This 
inhomogeneous field configuration can be stable on the quantum level and
provide a negative pressure component.
Ghost condensation and its relation to cosmological evolution has recently been investigated intensively \cite{Koe2015}, and
 kinematically driven acceleration of the Universe has been proposed in various frameworks \cite{Arm1999}. 

 Instead  of turning to some realistic dilaton or phantom field model of gravity, we shall here study the Euclidean, $O(2)$ symmetric scalar $\phi^4$ model, with modified kinetic energy term as a toy model, similar to that investigated in 
\cite{Lau2000}. The modification $-\Box \to \Box + Y\Box^2$ of the kinetic energy operator  would introduce  unboundedness
 of the Euclidean  action for $Y=0$, but for $Y>0$ it still remains bounded from below. Nevertheless the opposite signs of the quadratic and quartic gradient terms may result in occurring ghost condensation. For later convenience let us call scalar fields with $Z=1$ and $Z=-1$ ordinary and ghost fields, respectively.  
 In this paper we are going to use the Wegner-Houghton (WH) renormalization group (RG)
 equation \cite{Weg1973}  and below the singularity scale $k_c$ -- if there is any -- we deploy the tree-level renormalization (TLR), which is also called instability induced renormalization \cite{Ale1999}. The advantage of the WH RG scheme is the clearcut differentiation of handling the ultraviolet (UV) and the infrared (IR) modes of the field variable. Its main drawback is however that it is restricted to the local potential approximation (LPA), so that the gradient terms do not exhibit any RG flow. Thus in our present approach the dynamics providing the minimum of
the Euclidean action is governed by
the interplay of the bare gradient terms and the RG flow of the blocked
 local potential. 
 Functional RG schemes enabling one to go beyond the LPA
may reveal more dynamics due to the RG evolution of both the gradient terms and the local potential.

 As compared to the analysis in \cite{Lau2000}, our approach does not include the RG flow of the kinetic piece of the blocked action, but it takes with the flow of the full local potential including the quadratic mass term, the quartic as well as the higher-order polynomial terms. Therefore, it does not provide the possibility to replace the operator $\Omega(-\Box)$ by an effective one in the IR limit. On the other hand, we shall perform the TLR in  detail in order to determine the IR scaling laws even in  phases exhibiting spinodal instability, when those cannot be obtained from the WH RG equation. 
 
It is well-known that with the usual kinetic term the ordinary $O(N)$  models
in Euclidean space with the number of dimensions $d=3$ exhibit a symmetric phase
and a symmetry broken one, the scaling laws at the Gaussian, Wilson-Fisher, and IR fixed points serve as well-sounded test ground for any RG approach \cite{Tet1994}.   For $N=1$ the discrete $\phi\to -\phi$ symmetry, for $N>1$ the continuous $O(N)$ symmetry  is broken spontaneously by the nonvanishing homogeneous
vacuum field configuration and there occur $N-1$ Goldstone bosons. In the symmetric case the RG trajectories can be
followed up by means of the numerical solution of the WH RG equation, moving
the cutoff scale $k$ from the UV momentum cutoff $\Lambda$ down to the IR scale $k=0$. The IR limit of the blocked local potential keeps its polynomial form with the minimum at the field configuration $\phi=0$. In the symmetry broken phase there occurs a singularity of the logarithmic term of the WH RG equation at a finite momentum scale $k_c$ \cite{Ale1999}. This happens due to vanishing of the restoring force (the  term of the Euclidean action quadratic in the field variable) in the exponent of the integrand of the path integral. The system then starts to develop a spinodal instability. The WH RG equation loses its validity for scales  $k<k_c$ and the IR behaviour of the 
RG trajectories should be determined by the TLR procedure
taking explicitly into account the finite amplitude of the inhomogeneous mode
that minimizes the blocked action at the moving scale $k$. As a result the
 `Mexican hat' like local potential becomes convex in the IR limit reproducing the well-known Maxwell cut. Moreover, the approach to the singularity at the scales approaching $k_c$ from above can be revealed as an approach to an IR fixed point of the appropriately rescaled WH RG equations \cite{Nag2013,Tet1992}.

 In our case the modified kinetic energy operator $\Omega(-\Box)$ with $Z=-1$ 
may make the dynamics more rich. Namely, an inhomogeneous field configuration of finite amplitude may develop at a given scale $k<k_c$ even if no potential were present. The interplay of the gradient and the potential energy terms decides
the optimal amplitude $\rho_k$ of the inhomogeneous mode developed at the scale $k$. While for ordinary models with usual kinetic term a nonvanishing amplitude $\rho_k$ of the spinodal instability  developes due to the interplay of the positive kinetic term and the negative mass term of the blocked potential,
 for ghost models with  modified kinetic term $\Omega(-\Box)$ with $Z=-1$ the opposite signs of the various gradient terms may also responsible for developing a finite amplitude $\rho_k$ of the spinodal instability.  TLR enables one to study whether the amplitude $\rho_k$ of the inhomogeneous mode developed by the system  does survive the IR limit $k\to 0$ or not.
In our paper a particular emphasis is given to the determination of  the IR behaviour of the model by means of the TLR approach. In order to check our numerical procedure for TLR it has also been  applied to the symmetry broken phase of Euclidean one-component ordinary scalar field theory and  numerical results obtained supporting the theoretical analysis given in \cite{Ale1999}. As a further test, application of the numerical TLR procedure to the (ordinary) sine-Gordon model also reproduced the well-known IR behaviour in the molecular phase of the model \cite{Nan1999,Nag2007}.

It is another peculiarity of the $O(2)$ ghost scalar  model with the kinetic 
operator $\Omega(-\Box)$ that the coefficient $Y$ of the $\Box^2$ term is of natural dimension ${\rm{mass}}^{-2}$. Furthermore the coupling $Y$ is UV irrelevant, i.e.   a nonrenormalizable one. Nevertheless,  the ghost condensation when it takes place at some scale $k^2\approx (1/Y) $ plays a definitively decisive role in the IR physics of the model.    Therefore the coupling  $Y$ may become IR relevant.
The rather general way of distinguishing the various phases of the model is 
opened up by investigating the sensitivity of IR physics to the bare parameters of the model. Such kind of study is possible even if the dimensionful coupling $Y$ is kept constant at its bare value, like in the present work.
 Then the dimensionless coupling $\t{Y}=Yk^2$  scales down during the RG flow
 and
its interplay with the flow of the mass term of the potential  severely   
affects the scale at which the dimensionless inverse propagator may vanish.
In this approach no fixed points can be determined. Nevertheless, the global RG flow enables one to identify the phases determining their different IR
scaling behaviour and/or sensitivity to the bare parameters of the model. 
  The massive sine-Gordon model was successfully treated in a similar approach in \cite{Nag2008}. 

It should be mentioned that one could have chosen another approach when the
  the dimensionless coupling $\t{Y}$  would have been kept constant at its bare value during the RG flow. This would imply the blow up of the dimensionful coupling $Y=\t{Y} k^{-2}$ in the IR limit. Then the zero of the dimensionless inverse propagator might have been occur only due to the flow of the dimensionless mass parameter of the blocked potential. Such an approach, on the one hand, would have enabled  us 
to determine the fixed points of the model and to draw up the usual type of phase diagrams for various values of $\t{Y}$. On the other hand,  in that approach one would have loose the possibility to detect the effect of ghost condensation by decreasing the gliding cutoff $k$ downwards. The inverse propagator in momentum representation, $G^{-1}(k^2)=k^2(-1+\t{Y})+v_1(k)$ (with $v_1(k)$ the mass squared at the scale $k$) shows that the nonvanishing constant value of $\t{Y}$ is 
equivalent with a renormalization $Z=-1$ to $Z=-1+\t{Y}$. Therefore the
 eventual presence of the ghost condensation would have remained
 hidden by the RG procedure.

The paper is organized as follows. In Sec. \ref{sec:tests} we test our numerical
 TLR procedure applying it to the symmetry broken phase of the ordinary 3-dimensional Euclidean one-component scalar field model  and to the molecular phase of the 2-dimensional sine-Gordon model. In Sec. \ref{sec:O2} we identify the phases
of the  $O(2)$ symmetric ghost scalar model, determine the IR behaviour of its various phases, and determine the behaviour of the correlation length at the phase boundary. A comparison is also given to the ordinary counterpart of the model. Finally, the results are summarized in Sec. \ref{sec:sum}.

\section{One-component scalar field models}\label{sec:tests}

\subsection{Blocking transformation}

In the LPA the blocked action for the one-component scalar field $\phi(x)$
 in  3-dimensional Euclidean space has the form
\bea 
S_k[\phi]& =& \int d^3 x \frac{1}{2}\phi\Omega(-\Box) \phi + \int d^3 x U_k(\phi^2),
\eea
where $k$ is the gliding cutoff, $Z=1$ or $-1$, $Y\ge 0$, and $U_k(\phi)$ stands for the blocked potential. The blocking transformation corresponding to 
integrating out the modes of the field in the thin momentum shell $[k-\Delta k,k]$ is given as
\bea\label{block}
e^{-S_{k-\Delta k}[\phi]}&=&\int \mathcal{D}\phi' e^{-S_k[\phi+\phi']},
\eea
where the fields $\phi$ and $\phi'$ contain Fourier modes with momenta $p<k-\Delta k$, and $k-\Delta k<p<k $, respectively. (Throughout this paper we set $\hbar=1$ for the sake of simplicity.) This blocking transformation consists of integrating out the high-frequency Fourier components $\phi'$ of the field variable. In every infinitesimal step of the blocking given via Eq. \eq{block}, the integral is evaluated with the saddle point approximation. Two qualitatively different situations may occur depending on whether the second functional derivative of the blocked action {\em (i)} is positive definite or {\em (ii)} it starts to develop zero eigenvalues. In case {\em (i)} the saddle point is at $\phi'=0$ and  in the limit $\Delta k\to 0$ one arrives  to the WH equation
\bea\label{WH} 
k\partial_k U_k(\Phi)&=& -\alpha k^3\ln\biggl( \Omega(k^2)+\partial_\Phi^2U_k(\Phi) \biggr)
\eea 
in the LPA, where the field variable $\phi=\Phi=$const. is taken to be
 homogeneous, $\Omega(k^2)=Zk^2+Yk^4$ and $\alpha=\Omega_3/[2(2\pi)^3]=1/(4\pi^2)$ with the solid angle $\Omega_3=4\pi$.

In the case of unbroken $Z_2$ symmetry the RG trajectories can
 be followed up by means of Eq. \eq{WH} from the UV scale $k=\Lambda$ down to  the IR limit $k\to 0$. In the symmetry broken phase at some finite scale $k_c$ the situation {\em (ii)} appears. It is signalled by vanishing the argument of the logarithm in the right-hand side of Eq. \eq{WH}.  Now a non-trivial saddle point can be developed by the system, that minimizes the blocked action. Then Eq. \eq{WH} loses its validity and one has to turn to the TLR procedure and rewrite the blocking relation \eq{block} into the form
\bea\label{block2}
  S_{k-\Delta k} [\phi] &=& {\mbox{min}}_{\phi'} S_k [\phi+\phi'],
\eea
where the loop integral has been neglected completely. Reducing the functional
space of the saddle-point configurations $\phi'$ to those of periodic ones of the form
\bea\label{plwa} 
\psi_k(x)&= &2\rho\cos(k n_\mu(k)x_\mu+\theta(k)),
\eea 
the action $ S_k [\phi+\psi_k]$ becomes a function of the amplitude $\rho$.
 Here $n_\mu(k)$ and $\theta(k)$ are
a spatial unit vector and a phase shift, respectively. 
Let $\rho_k$ be the value of the amplitude of the saddle-point configuration
for which the action $ S_k [\phi+\psi_k]$ takes its minimum value. 
 Various saddle points
 of the system corresponding to various values of $n_\mu(k)$ and $\theta(k)$ 
are physically inequivalent but are expected to belong to the same minimal value of the blocked action. Inserting  ansatz \eq{plwa} into the tree-level blocking relation \eq{block2} one finds
\bea\label{blocktree} 
U_{k-\Delta k}(\Phi)&=&{\mbox{min}}_{\{\rho\}}\biggl(\Omega(k^2)\rho^2\nn
&&+\frac{1}{2}\int_{-1}^1 du U_k \bigl( \Phi+2\rho\cos(\pi u)\bigr) \biggr).
\eea
Due to spatial $O(3)$ symmetry  the expression in the braces in the
 right-hand side of Eq. \eq{blocktree} only depends on $\rho$, as expected.

\subsection{Polynomial potential}\label{polypo}

For the local potential chosen  in the Taylor-expanded form
\bea\label{polpot}
 U_k(\Phi)&=&\sum_{n=0}^M\frac{v_n}{n!} r^{n}= \sum_{n=0}^M \frac{g_{2n}}{(2n)!}
\Phi^{2n}
\eea
with $r=\hf\Phi^2$, $g_{2n}=\frac{ (2n)!}{n!2^n} v_n$ and truncated at $n=M$ the WH RG equation \eq{WH} can be rewritten as a set of coupled ordinary, first order differential equations for the dimensionless running couplings $\t{v}_n(k)$. For $M=2$ those are
\bea\label{whv1v2} 
k\partial_k \t{v}_1&=&-a\alpha\frac{\t{v}_2}{\t{v}_1+Z+Yk^2}-2\t{v}_1,\nn
k\partial_k \t{v}_2&=&b\alpha\frac{\t{v}_2^2}{(\t{v}_1+Z+Yk^2)^2}-\t{v}_2
\eea
with $a=1$ and $b=3$.  The dimensionless couplings $\t{v}_1$ and $\t{v}_2$ are defined by $v_1=k^2\t{v}_1$ and $v_2=k\t{v}_2$. Since   the phase structure and the scaling laws in the various scaling regimes do not alter qualitatively with increasing truncation $M$, we shall work with $M=2$ when solving the WH RG equations and choose $\Lambda=1$ throughout this work.

As discussed above the RG trajectories belonging to the symmetry broken phase
can be followed  by the WH RG equation down to the scale $k_c$ at which the
right-hand side of Eq. \eq{WH} becomes singular. The  IR scaling can, however,  be determined by means of the TLR procedure, following the RG trajectories below the critical scale $k_c$ down to the IR limit $k\to 0$.   The  TLR
 procedure discussed in more detail  in Ref. \cite{Ale1999} 
 is shortly summarized in Appendix \ref{treelerg}. The same TLR procedure can be extended for ghost models with kinetic energy operator $\Omega(-\Box)$   in a straighforward manner as follows. For scales $k<k_c$ spinodal instability occurs when the logarithm in the right-hand side of Eq.  \eq{WH} satisfies
 the inequality  
\bea\label{singcond}
 Z+Yk_c^2+\t{v}_1(k_c)+\frac{3}{2}\t{v}_2(k_c)\t{\Phi}^2&\le &0.
\eea
Since the last term in the left-hand side of the inequality is positive,
the singularity occurs at $\Phi=0$ with decreasing scale $k$ when the condition 
\bea
  Z+Yk_c^2+\t{v}_1(k_c)&=&0
\eea
is met.
For $Z=+1$ and $Y=0$, this yields $1+\t{v}_1(k_c)=0$, and generally there exists such a scale $k_c$ in the symmetry broken phase; the critical scale is governed by the negative (dimensionless) mass squared in the potential. For $Z=-1$, $Y=0$ we find the
condition for occuring the singularity when $-1+\t{v}_1(k_c)=0$, now with 
positive mass term of the potential.  For $Z=-1$, $Y>0$ the condition for
occurring the singularity becomes
\bea
  -1 +\t{v}_1(k_c)+ Yk_c^2&=&0,
\eea
and means that an interplay of the quartic gradient term and the mass term
determines the scale $k_c$. Supposing that it holds $\t{v}_1(k_c)<0$, 
 the critical scale is $k_c^2= [1-\t{v}_1(k_c)]/Y$ and for $|\t{v}_1(k_c)|\ll 1$
one finds $k_c^2\sim  \ord{1/Y}$. In such cases  ghost condensation in the modes with $k<k_c$  takes place and may play a decisive role in the behaviour of the phase in the deep IR region.  
The equality in \eq{singcond} with $k_c$ replaced by the moving cutoff $k<k_c$
determines a critical field amplitude $\Phi_c(k)=\sqrt{k}\t{\Phi}_c(k)$ such that an interval
 $|\Phi|< \Phi_c(k)$ opens up with decreasing scale $k$ in which spinodal instability occurs.
For $k<k_c$ the critical field amplitude is determined by the equality
\bea
 Z+Yk^2+\t{v}_1(k)+\frac{3}{2}\t{v}_2(k)\t{\Phi}^2_c&=&0
\eea
as
\bea\label{phick1sc}
\t{\Phi}_c(k)&=& \sqrt{ 2\lbrack-Z-Yk^2-\t{v}_1(k)\rbrack/3\t{v}_2(k)}.
\eea
The interval $|\Phi|\le \Phi_c(k)$ of instability survives the limit $k\to 0$
if and only if $\sqrt{k}\t{\Phi}_c(k) $  takes a finite or infinite limit which restricts the IR scalings of the couplings  $\t{v}_1(k)$ and $\t{v}_2(k)$.

For scales $k<k_c$ and background fields $\Phi\in [-\Phi_c,\Phi_c]$ one turns to the tree-level blocking relation \eq{blocktree} and inserting the ansatz \eq{polpot} into it, one obtains the recursion relation 
\bea\label{blocktree2k2}
U_{k-\Delta k}(\Phi)&=&\min_{\{\rho\}}\biggl(U_k(\Phi)+(Z+ Yk^2)k^2\rho^2\nn
&&+\sum_{n=1}^M \frac{\rho^{2n}}{(n!)^2}\partial_\Phi^{2n} U_k(\Phi)\biggr)
\eea 
for the running couplings \cite{Ale1999}. For given scale $k$ with given couplings $v_n(k)$ and
for given homogeneous field $\Phi\in [-\Phi_c,\Phi_c]$, one determines the value $\rho_k(\Phi)$ minimizing the right-hand side of Eq.
\eq{blocktree2k2}. Then one repeats this minimization for various $\Phi$ values and determines the corresponding $U_{k-\Delta k}(\Phi)$ values. Finally these discrete values of   $U_{k-\Delta k}(\Phi)$ are fitted 
by the polynomial \eq{polpot} in the interval $\Phi\in [-\Phi_c,\Phi_c]$ in order to read off the new $v_n(k-\Delta k)$ of the couplings. In such a manner the behaviour of the RG trajectories can be investigated in the deep IR region. This numerical procedure generally converges
for sufficiently small values of the ratio $\Delta k/k$.
The blocked potential $U_{k<k_c}(\Phi) $  outside of the interval $-\Phi_c\le \Phi \le \Phi_c$ can be taken identical to $U_{k_c}(\Phi)$ with good accuracy, because there it suffers no tree-level renormalization \cite{Ale1999}.
In Sec. \ref{sec:o2whrg} we shall argue that the TLR of the ghost scalar field with $O(2)$ symmetry can be reduced to the case of the TLR of the real one-component ghost scalar field when the nontrivial saddle-point configuration is looked for in an appropriately reduced functional space. Numerical study of that case shall be pursued in Sec. \ref{sec:o2phdi}. 

Here we concentrate on the test of our numerical procedure for TLR, applying it to the 3-dimensional Euclidean polynomial model of the ordinary one-component scalar field with $Z_2$ symmetry, considering the case with  $Z=1$, $Y=0$. Choosing the truncation of the polynomial potential at $M=10$, we achieved good numerical convergence   for $\Delta k/k=0.001$ and the least square fitting procedure with the number $60$ of equidistant  grid points in the interval  $-\Phi_c(k)\le \Phi \le \Phi_c(k)$. It has been checked that the results are stable against increasing the number of grid points.  In order to achieve a better least square fit, the  TLR procedure has been performed in the wider interval $\Phi\in [-\b{\Phi},\b{\Phi}]$ with  $\b{\Phi}=\sqrt{ -2v_1(k_c) /3v_2(k_c)} $. The latter is a good estimate of $\Phi_c(k)$ for $k\ll k_c$ \cite{Ale1999}. It has been established numerically that the blocked potential does not acquire any tree-level correction outside of the interval of instability $\Phi\in[-\Phi_c(k),\Phi_c(k)]$.

\begin{figure}[ht]
\centerline{\epsfig{file=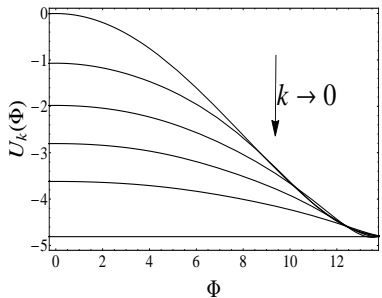,height=4.0cm,width=5.0cm,angle=0}}
\caption{\label{potflow}
The blocked potential at various scales $k$. }
\end{figure}

\begin{figure}[ht]
\centerline{\epsfig{file=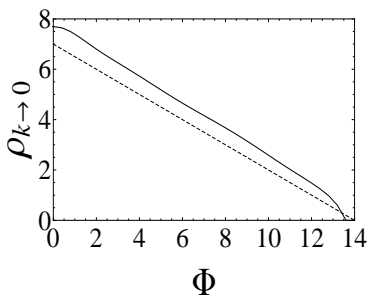,height=4.0cm,width=5.0cm,angle=0}}
\caption{\label{rho}
Our numerical result for  $\rho_k(\Phi)$ in the IR limit $k\to 0$ (solid line)
in comparison with the one reported in \cite{Ale1999} (dashed line).
}
\end{figure}

\begin{figure}[ht]
\centerline{\epsfig{file=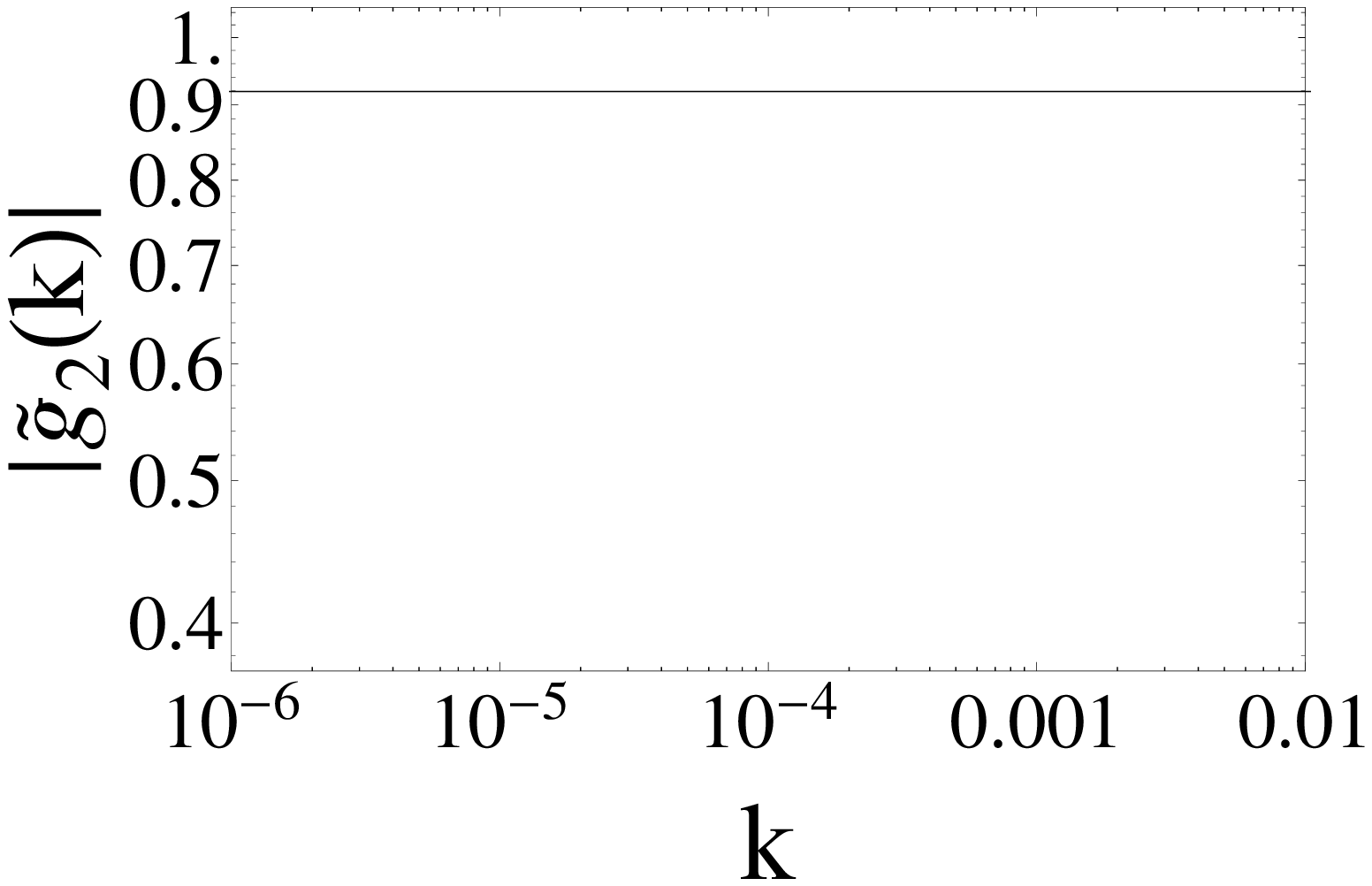,height=3.20cm,width=4.00cm,angle=0}\epsfig{file=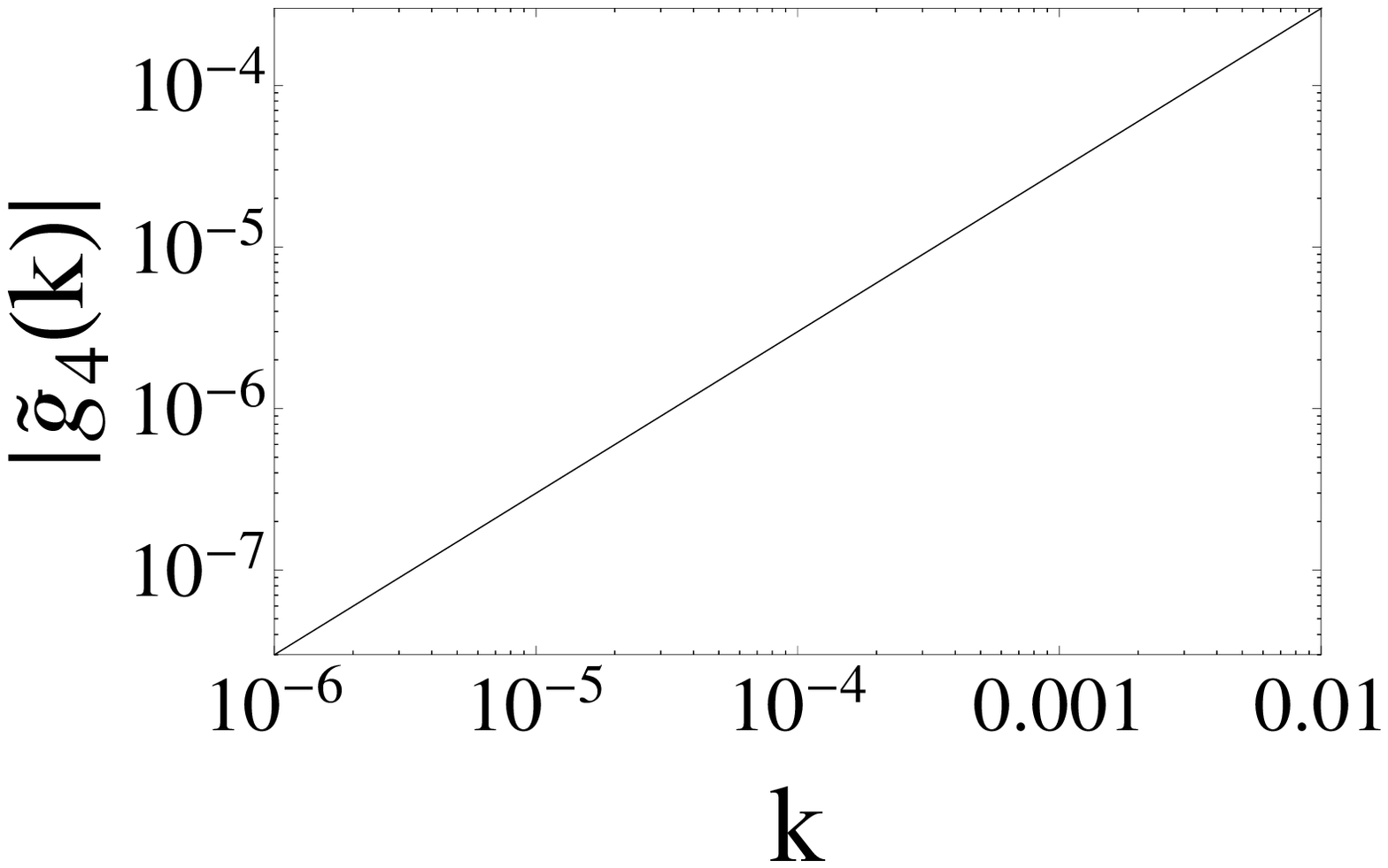,height=3.20cm,width=4.00cm,angle=0}}
\centerline{\epsfig{file=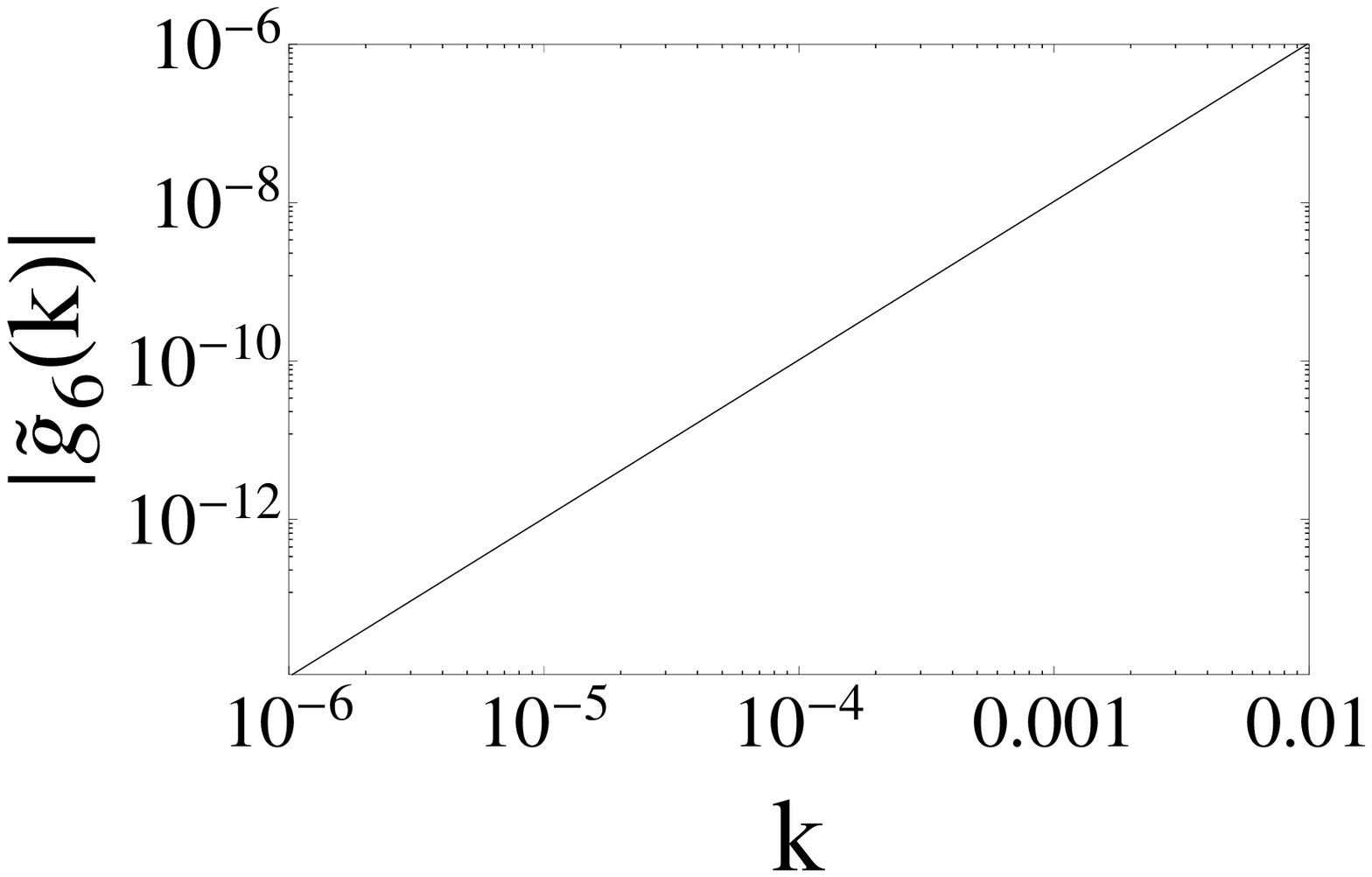,height=3.20cm,width=4.00cm,angle=0}}
\caption{\label{cscaling}
Scaling of the dimensionless couplings $\t{g}_2=\t{v}_1$, $\t{g}_4=3\t{v}_2$ and $\t{g}_6=15\t{v}_3$.}
\end{figure}

\begin{figure}[ht]
\centerline{\epsfig{file=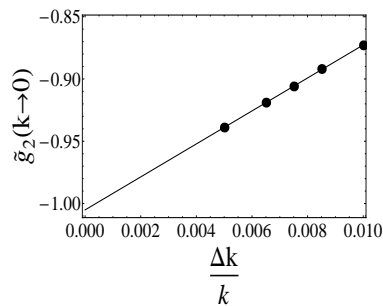,height=4.0cm,width=5.0cm,angle=0}}
\caption{\label{extp}
Determining the value $\t{g}_2=\t{v}_1$ in the limit $k\to 0$   at effectively zero step size from extrapolation.}
\end{figure}

According to our numerical results shown in Fig.  \ref{potflow}, the dimensionful blocked potential $U_k(\Phi)$ tends to and reaches the Maxwell-construction in the  limit $k\to 0$, as expected \cite{Ale1999}.
For the scale $k\approx 10^{-6}$ the function  $\rho_k(\Phi)$ obtained numerically is shown in Fig. \ref{rho}  in comparison with the  curve $\rho_k(\Phi)= (-\Phi+\Phi_c)/2$ obtained in Ref. \cite{Ale1999}. The slope $-0.53$ obtained numerically is in good agreement with its theoretical value $-0.5$.
 It should be emphasized that the dimensionful amplitude $\rho_k$ of the spinodal instability survives the IR limit with $2\rho_{k\to 0}(\Phi=0)\approx \Phi_c$, so that
on vanishing background $\Phi=0$ the instability pushes the field configuration
to the homogeneous one at either $\psi_{k\to 0}=2\rho_{k\to 0}(0)= \Phi_c$ or $-\Phi_c$, both of them belonging to the same constant value of the effective potential.  
Also the IR scaling of the couplings has been established. On the one hand, 
we have  determined the scaling of the dimensionless couplings $\t{v}_1+1 \sim k^{\alpha_1}$, $\t{v}_2 \sim k^{\alpha_2}$, and $\t{v}_3 \sim k^{\alpha_3}$ in the terms $\Phi^2$, $\Phi^4$, and $\Phi^6$, respectively (see Fig. \ref{cscaling}) and obtained the exponents $\alpha_1 = 0.08\pm 0.08$, $\alpha_2= 1\pm 0.001$, and $\alpha_3=1.34\pm 0.01$. The errors of $\alpha_2$ and $\alpha_3$ are those of the log-log fit.  It should be noticed, however, that the run of $1+ \t{v}_1(k)$ is extremely slowed down in the deep IR region, so that the
 numerical determination of the exponent $\alpha_1$ may have an error comparable to its magnitude.
 On the other 
hand, numerics revealed without any doubt that $\Phi_c(k)\sim \sqrt{ k k^{\alpha_1-\alpha_2}}$ is finite, providing the restriction that the equality $1+\alpha_1-\alpha_2=0$ should hold implying that $\alpha_1\approx 0$ with high accuracy. 
It is known that $\t{v}_1(k)\to -1$ and $\t{v}_{n>1}\to 0$ in the IR limit $k\to 0$ and that limit corresponds to the   RG invariant  effective potential $\t{U}_{k\to 0}(\t{\Phi}) = -\hf \t{\Phi}^2$
 in the interval $\lbrack -\t{\Phi}_c, \t{\Phi}_c\rbrack$ \cite{Ale1999}.
In our numerical calculations, $\t{v}_1(k)$ tends to a constant value close to $-1$, but this value turned out to  decrease linearly with decreasing step size $\Delta k/k$. In order to fix the IR limiting value of $\t{v}_1$ numerically, we calculated it for five different step sizes, and the extrapolation to $\Delta k /k \approx 0$,  shown in Fig. \ref{extp},  yielded the extrapolated value $\t{v}_1^{ext}(0)=-1.005$.
We conclude that our numerical results for the IR behaviour of the ordinary one-component
scalar field completely reproduce those reported and argued for in Ref. \cite{Ale1999}.

\subsection{Tree-level renormalization of the sine-Gordon model}

Another verification of our numerical apparatus for TLR
has been obtained from its  application
 to the 2-dimensional
Euclidean sine-Gordon model given by the classical action
\bea 
S[\phi]&=&\int d^2 x  \biggl[\frac{1}{2}(\partial_\mu \phi)^2+ u_1 \cos(\beta \phi) \biggr].
\eea
 The parameter region $\beta^2 < 8\pi$ belongs to the spontaneously broken phase of the model, while for $\beta^2 > 8\pi$ we can find the symmetric phase.
The results of the TLR of the sine-Gordon model are well known
 \cite{Nag2013,Nag2007,Nan1999} and provide another test of our numerical
 procedure. The tree-level blocking relation \eq{blocktree} for the ansatz
\bea 
U_k(\Phi)&=&\sum_{n=0}^M u_n(k)\cos(n\beta \Phi)
\eea
 can be rewritten now in the form of the recursion equation 
\bea 
U_{k-\Delta k}(\Phi)&=& {\mbox{min}}_{\rho}\biggl[k^2\rho^2+\sum_{n=0}^M u_n(k)\cos(n\beta \phi)J_0(2n\beta \rho) \biggr]\nn
\eea
(see Ref. \cite{Nan1999}).
Here $J_0$ stands for the Bessel function, and the potential is truncated at the $M$-th upper harmonic.

\begin{figure}[ht]
\centerline{\epsfig{file=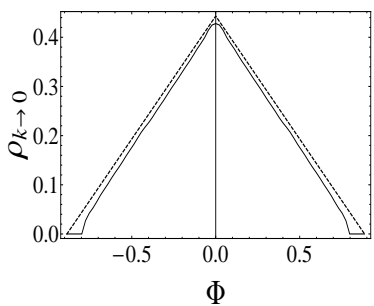,height=4.0cm,width=5.0cm,angle=0}}
\caption{\label{sgrho} 
Comparison of the function
$\rho_k(\Phi)$ obtained by us numerically (solid line) to the one given in Ref. \cite{Nan1999} (dashed line) for  the molecular phase of the SG model for $\beta^2=4\pi$.
}
\end{figure}

\begin{figure}[ht]
\centerline{\epsfig{file=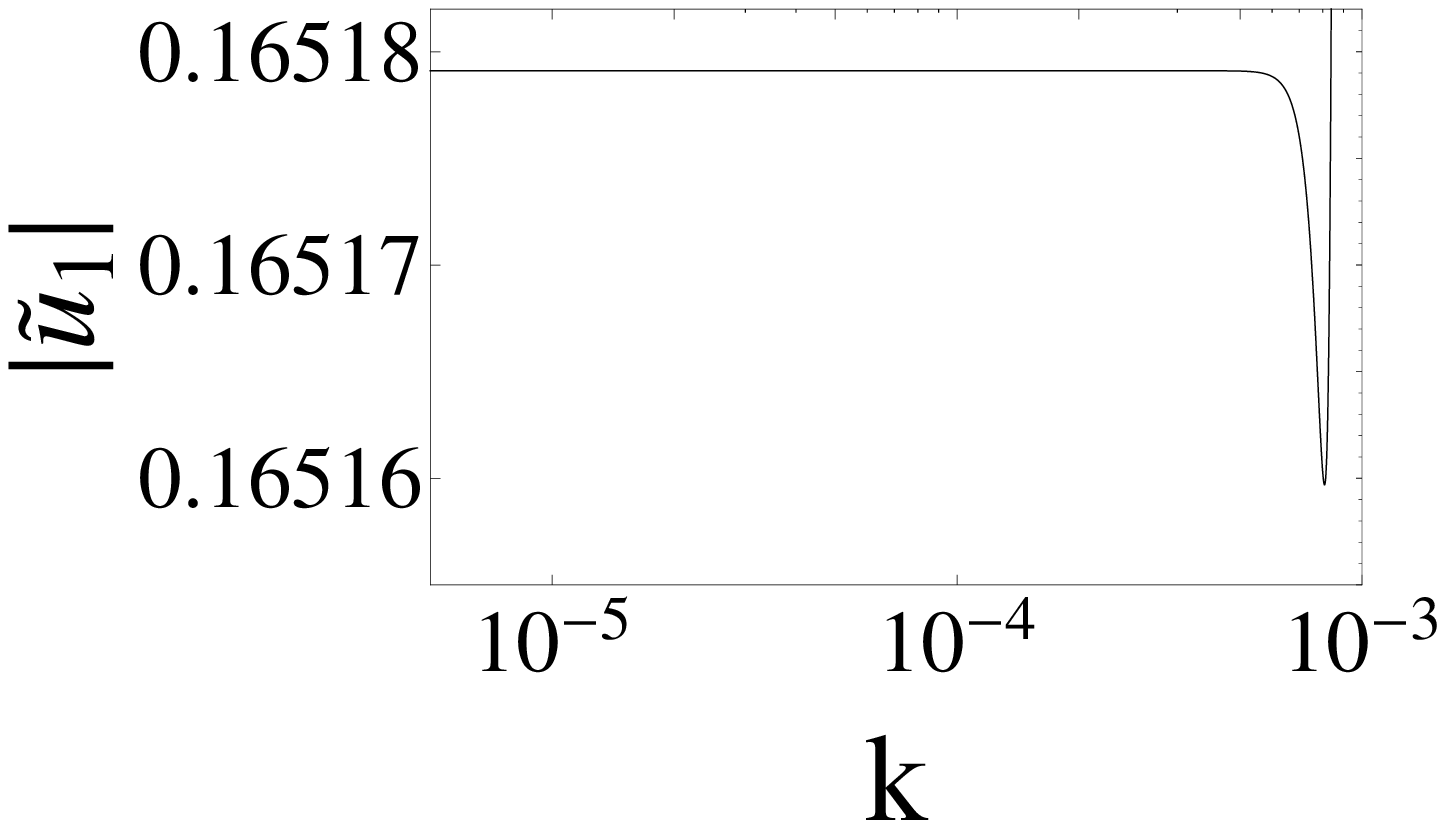,height=3.52cm,width=4.4cm,angle=0}\epsfig{file=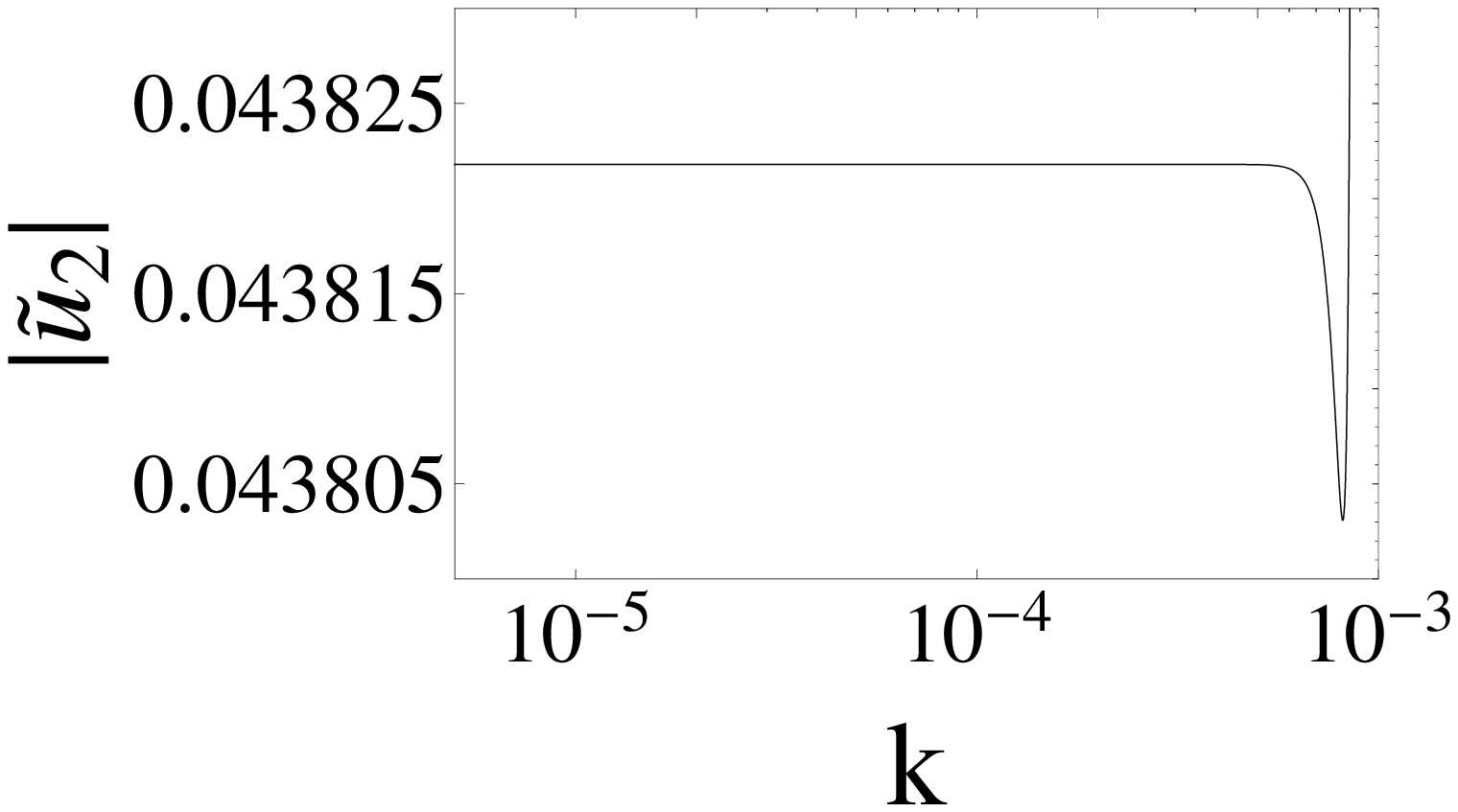,height=3.52cm,width=4.4cm,angle=0}}
\centerline{\epsfig{file=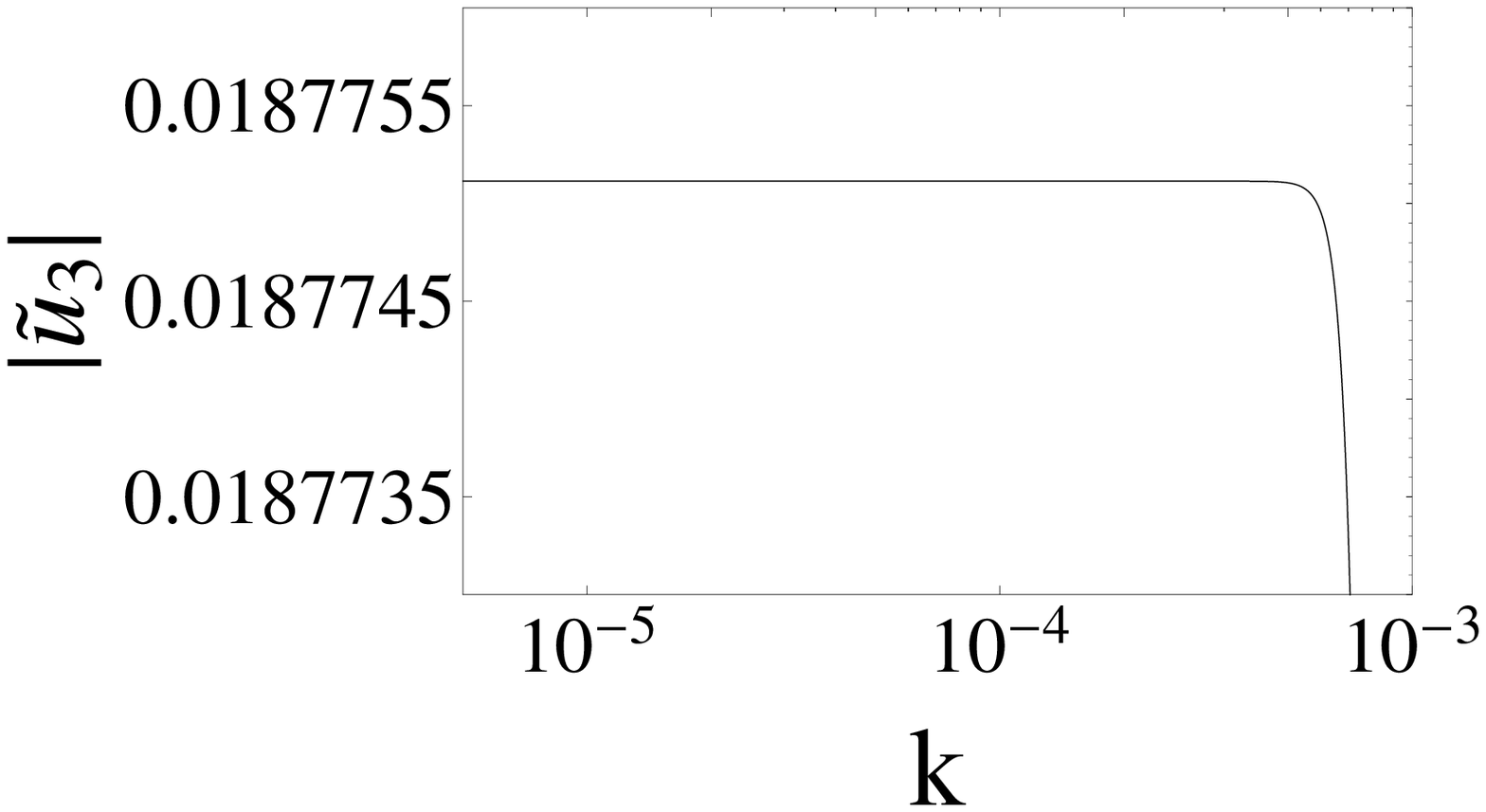,height=3.52cm,width=4.4cm,angle=0}\epsfig{file=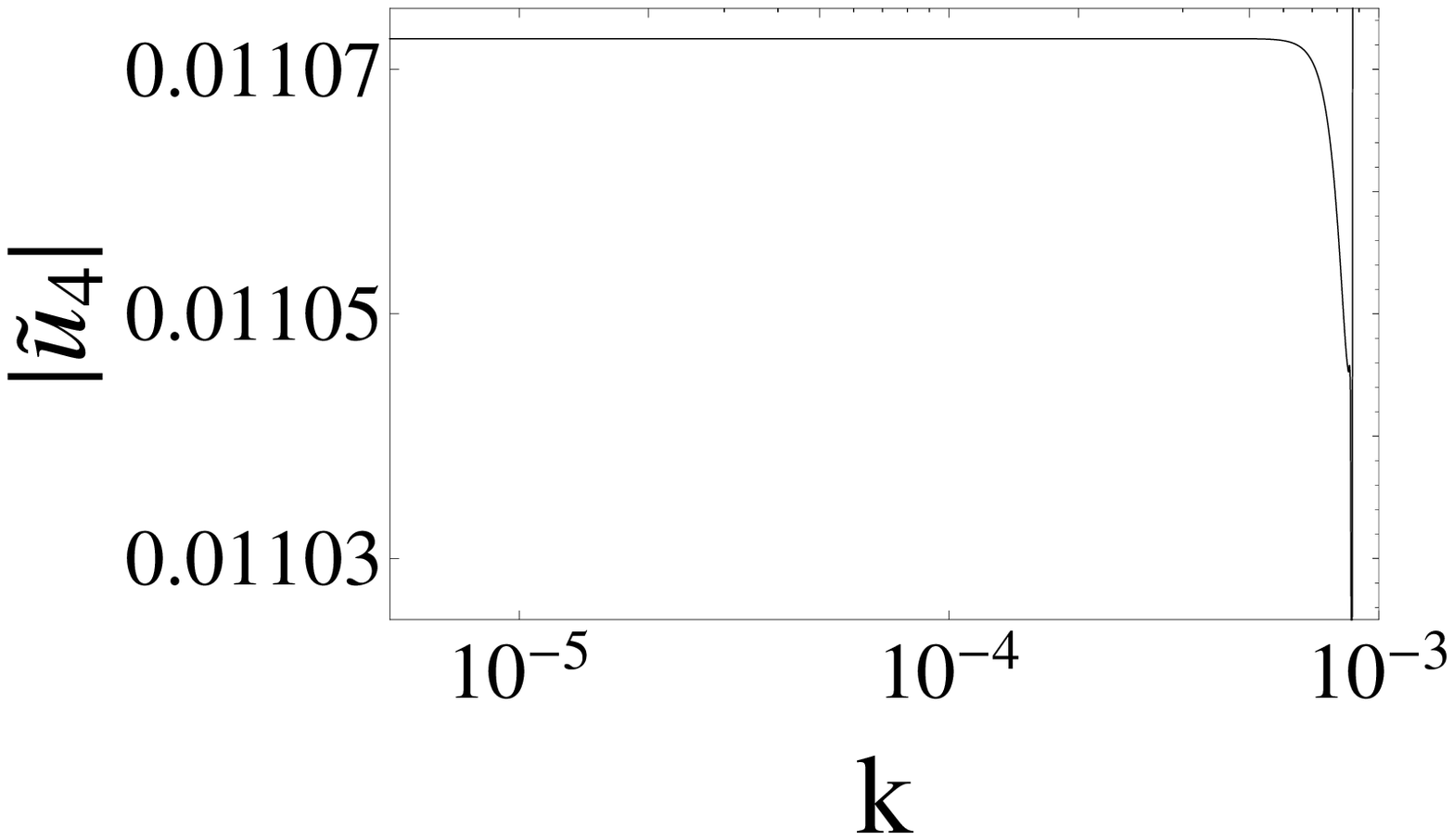,height=3.52cm,width=4.4cm,angle=0}}
\caption{\label{sgcouplings}
The scale-dependence of the dimensionless running couplings of the SG model in the deep IR region for 
 $\beta^2=4\pi$}
\end{figure}

 For the numerical calculations we set  $\beta^2=4\pi$ and  $M=10$.
The outcome of our numerical calculations is in complete agreement with 
the literature. Under the scale  $k_c$, where the spinodal instability occurs, it is known  that the amplitude $\rho(\Phi)$ of the periodic field configurations, which minimizes the action, is given by $\rho_k(\Phi)=-\frac{1}{2}(|\Phi|-\frac{2\pi}{\beta})$ \cite{Nan1999}.
Similarly to the spontaneously broken phase of the one-component scalar field theory with polynomial interaction, the amplitude of the spinodal instability survives the IR limit again.
 The comparision of our numerical result for $\rho_k(\Phi)$ to   the
 one obtained in Ref. \cite{Nan1999} can be seen in Fig. \ref{sgrho}.
We also plotted in Fig. \ref{sgcouplings} the magnitude of the first four dimensionless couplings, which in fact, are renormalizable and tend to a constant value in the $k \to 0$ limit. This means that the dimensionful effective potential
becomes vanishing in accordance with the requirements of convexity and periodicity \cite{Nan1999}.
Thus we conclude that our numerical procedure applied for the TLR yields results for the IR behaviour of the sine-Gordon model in complete agreement with those obtained in Refs. \cite{Nag2013,Nag2007,Nan1999}.

\section{$O(2)$ scalar model}\label{sec:O2}

\subsection{Application of the WH RG approach}\label{sec:o2whrg}

Let us turn now to our main goal, the study of the 3-dimensional,
 Euclidean, $O(2)$ symmetric model for the ghost scalar field $\phi$ with polynomial potential, using the LPA ansatz
\bea \label{realaction}
S_k[\underline{\phi}]& =& \frac{1}{2}\int d^3 x \underline{\phi}^T
\Omega(-\Box )\underline{\phi} + \int d^3 x U_k( \u{\phi}^T\u{\phi} ),
\eea
for the blocked action,
where  $\underline{\phi}
=\begin{pmatrix}\phi_1\cr\phi_2\end{pmatrix}$ denotes  the two-component real scalar field and
$U_k( \u{\phi}^T\u{\phi} )$ stands for the blocked potential.
For the latter we shall use the ansatz \eq{polpot} with $r=\u{\phi}^T\u{\phi}$.
 For comparison, we shall also discuss the behaviour of the model for ordinary scalar field with $Z=1$.

 In order to determine the phase diagram of  the $O(2)$ symmetric scalar ghost model, we apply the WH RG method again.
In Appendix \ref{app:o2whrg} we derive the WH RG equation
\bea\label{WHO2} \!\!\!\!\! \!\!\!\!
k\partial_k U_k(r)&=&-\alpha k^3\biggl\lbrack
 \ln \lbrack \Omega(k^2)+ U_k'(r)+2r U_k''(r)\rbrack \nn
&&+
\ln \lbrack \Omega(k^2)+ U_k'(r)\rbrack
 \biggr\rbrack
\eea
with $U_k'(r)=\partial_r U_k(r)$ and $U_k''(r)=\partial_r^2 U_k(r)$.
Eq. \eq{WHO2} for $Z=1$, $Y=0$ is just the particular case $N=2$ of the WH equation
 for $O(N)$ symmetric $\phi^4$ models,
\bea\label{litwh}
k\partial_k U_k(\Phi)&=&-\alpha k^3\biggl\lbrack \ln \lbrack k^2+\partial_\Phi^2U_k(\Phi) \rbrack
\nn&&+
\ln \lbrack k^2+\frac{1}{\Phi}\partial_\Phi U_k(\Phi) \rbrack^{N-1} 
\biggr\rbrack,
\eea
given in Ref. \cite{Ale1999}. To reveal the complete agreement of  our result with Eq. \eq{litwh} for $N=2$, one  has to make the substitution $\Phi=\sqrt{2r}$.

It is trivial that the $U(1)$ symmetric ansatz
\bea\label{skomplex} 
S_k[\phi^*,\phi] &=& \int d^d x \phi^*\Omega(-\Box )\phi + \int d^d x U_k(\phi^*\phi)
\eea
for the blocked action of the one-component complex  scalar field $\phi = \frac{1}{\sqrt{2}} (\phi_1+i \phi_2)$ is equivalent with the ansatz \eq{realaction}.
 In Appendix \ref{app:u1whrg} it is  shown that both of the  blocked actions given
by Eqs. \eq{realaction} and  \eq{skomplex}  yield  the same WH RG 
 equation for the blocked potential. In order to avoid numerical work with complex numbers,  we shall apply numerically the WH RG scheme to the $O(2)$ symmetric case.

The applicability of the WH RG equation may break down at some
 scale $k_c$ because the argument of the logarithm on the right-hand side
 of Eq. \eq{WHO2} can eventually reach zero. This occurs, when either of the conditions $s_-(k)=\biggl[\Omega(k^2)+U_k'(r)\biggr]\leq 0$ or $s_+(k)=\biggl[\Omega(k^2)+ U_k'(r)+2r U_k''(r)\biggr]\leq 0$ is fulfilled \cite{Ale1999}. This is the case of a spontaneously broken symmetry.
 These conditions mean, that the loop expansion is inapplicable when $k\leq k_c$. The  expression $s_-(k)$ is the inverse propagator of the lightest excitations of the field, the Goldstone-bosons. In the $O(N)$ symmetric model with a homogeneous vacuum field configuration pointing into a given direction of the internal space, there are $N-1$ transversal excitations or Goldstone-bosons, as it can be seen from the power $N-1$ of the eigenvalue $s_-(k)$ under the logarithm in the right-hand side of Eq. \eq{litwh}.
As mentioned before, in the case of the one-component scalar field, i.e., in the case with $N=1$, the vanishing of $s_+(k)$ drives the occurence of spinodal instability. For $N\ge 2$ the vanishing of $s_-(k)$ takes over that role. The critical scale $k_c$ is given by $s_-(k_c)|_{\Phi=0}=0$ implying
 $Z+Yk_c^2+\t{v}_1(k_c)=0$, just like in the case  $N=1$. For local potentials monotonically increasing for asymptotically large values of $|\Phi|$ and for scales $k<k_c$
 the interval $0\le |\Phi|\le \Phi_c(k)$ (with $\Phi_c(k)=\sqrt{k}\t{\Phi}_c(k)$) of instability may open up determined via the vanishing of $s_-(k)$  
as
\bea\label{phico2}
  \t{\Phi}_c(k) &=& 
 \sqrt{ - \frac{ 2  \lbrack Z+Yk^2 +\t{v}_1(k) \rbrack }{
  \t{v}_2(k) } } .
\eea

 Supposing that a nontrivial
saddle point $\u{\phi}'=\u{\psi}_k$ appears in the integrand  on the right-hand side of Eq. \eq{blockO2}, the integral can be approximated by the contribution of that saddle point. Thus one finds the relation, the generalization of Eq. \eq{block2}, 
\bea\label{block2O2}
  S_{k-\Delta k}[\u{\phi}] &=& {\mbox{min}}_{\{\u{\phi}'\}} S_{k}[
   \u{\phi}+\u{\phi}']=  S_{k}[
   \u{\phi}+\u{\psi}_k],
\eea
where $\u{\psi}_k(x)\not=0$ represents the nontrivial  saddle-point configuration
minimizing the action $ S_{k}[   \u{\phi}+\u{\phi}']$. For practical purpose 
we restrict ourselves to looking for nontrivial saddle-point configurations in a particular subspace of the configuration space, say to periodic configurations of the type given in  Eq. \eq{plwa}.

In the $O(N)$ case there are,
 however, possibilities to choose the nontrivial saddle-point configuration with various orientations in the internal space. Being restricted to LPA by the WH approach, the background configuration should be chosen homogeneous,
$\u{\phi}=\u{\Phi}=\Phi\u{e}$ pointing to some particular direction given by the unit vector $\u{e}$ in the internal space. In general, the nontrivial saddle-point configuration might have components parallel and orthogonal to the direction $\u{e}$. The question arises how these components should be chosen in order to 
minimize the value of the action. 
It was argued in Ref. \cite{Ale1999} that the TLR of ordinary $O(N)$ models for $N\ge2$  can be reduced to the TLR of the ordinary one-component  scalar model. The argument is based on the positivity of the quadratic gradient term. Without loss of generality, the field configuration $\u{\Phi}+\u{\psi}_k$
can be rewritten as
\bea\label{consad} 
\Phi \u{e}+\u{\psi}_k(x)&=&\eta_k(x)\u{\u{\c{R}}}(x)\u{e},
\eea
in terms of an appropriately chosen amplitude function $\eta_k(x)$ with the
$SO(N)$ matrix $\u{\u{\c{R}}}(x)$. Now the quadratic gradient term of the action
takes the form
\bea\label{quadgrad}
\lefteqn{
\hf \int d^d x \biggl( \partial_\mu \lbrack \eta_k(x)\u{\u{\c{R}}}(x)\u{e}\biggr)^T
   \biggl( \partial_\mu\lbrack \eta_k(x)\u{\u{\c{R}}}(x)\u{e}\rbrack\biggr)  }
~~~~~~~~\nn
&=&
 \hf \int d^d x \biggl( \lbrack \partial_\mu \eta_k(x)\rbrack \lbrack \partial_\mu \eta_k(x)\rbrack \nn
&&+ \eta_k^2(x) \lbrack \partial_\mu\u{\u{\c{R}}}(x)\u{e}\rbrack^T
\lbrack \partial_\mu\u{\u{\c{R}}}(x)\u{e}\rbrack \biggr),
\eea
where the identities $[\u{\u{\c{R}}}(x)\u{e}]^T\u{\u{\c{R}}}(x)\u{e}=1$ and 
$[\u{\u{\c{R}}}(x)\u{e}]^T
\partial_\mu [\u{\u{\c{R}}}(x)\u{e}]=0$  have been used. This means that any
inhomogeneity of the vector $\u{\u{\c{R}}}(x)\u{e}$ yields a positive contribution to the action, so that the nontrivial saddle point should be such that $\u{\u{\c{R}}}(x)\u{e}$  were homogeneous.  Then the relation \eq{consad} implies that both of the vectors $\u{\u{\c{R}}}(x)\u{e}$ and $\u{\psi}_k(x)$ should be parallel to the direction 
$\u{e}$ of the background field. The periodic ansatz 
for the nontrivial saddle-point configuration, similar to the one given by Eq. 
\eq{plwa}, is then
\bea\label{plwaO2}
  \u{\psi}_k (x)&=& \u{e} 2\rho_k \cos (kn_\mu(k) x_\mu+\theta_k).
\eea
Inserting it into the tree-level blocking relation \eq{block2O2} one arrives  at Eq.  \eq{blocktree} that can be recasted in the form of the recursion relation  \eq{blocktree2k2}.
So the TLR procedure of the ordinary $O(N)$ model reduces to the that for the ordinary $O(1)$ model.

For $Z=-1$, i.e., for the $O(N)$ ghost models with negative quadratic gradient term
the  above given argumentation fails because the terms in Eq. \eq{quadgrad}
acquire negative signs and no conclusion can be made that $\c{R}\u{e} $ were 
homogeneous. Here we shall make the ansatz \eq{plwaO2} for the nontrivial saddle point again. It might happen however that similar periodic saddle-point configurations with more sophisticated orientation in the internal space could give
smaller value of the blocked action. When the ansatz \eq{plwaO2} is used,
  the TLR of the $O(2)$ ghost model reduces to the TLR of the polynomial model of the one-component scalar field, except that the interval of constant background fields in which the spinodal instability occurs is now determined by the critical value $\t{\Phi}_c(k)$ given in Eq. \eq{phico2} instead of Eq. \eq{phicone}. The tree-level blocking relation \eq{block2O2} results in the recursion equation \eq{blocktree2k2} for the blocked potential, again.

\subsection{Phase structure and IR scaling laws}\label{sec:o2phdi}

\subsubsection{Identification of the phases}
\begin{figure}[t]
\centerline{\psfig{file=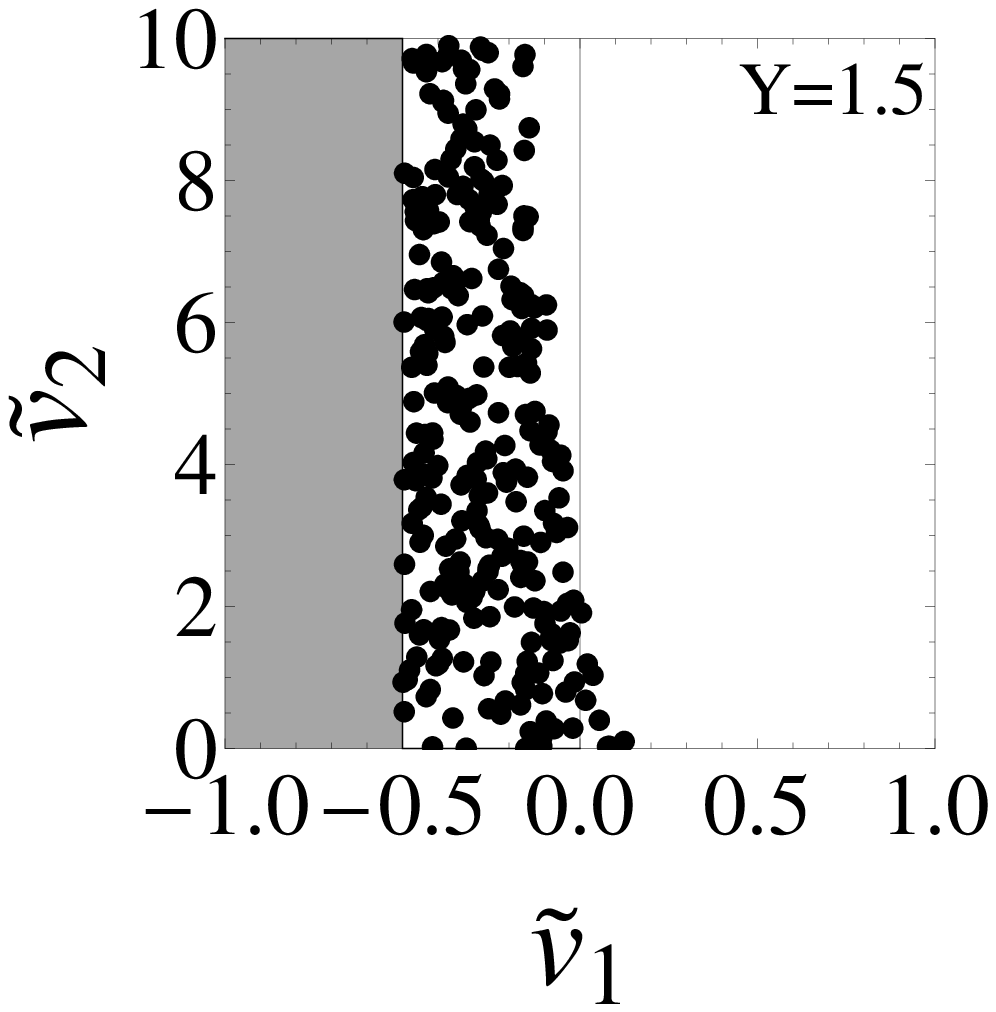,height=3.52cm,width=4.4cm,angle=0}
\psfig{file=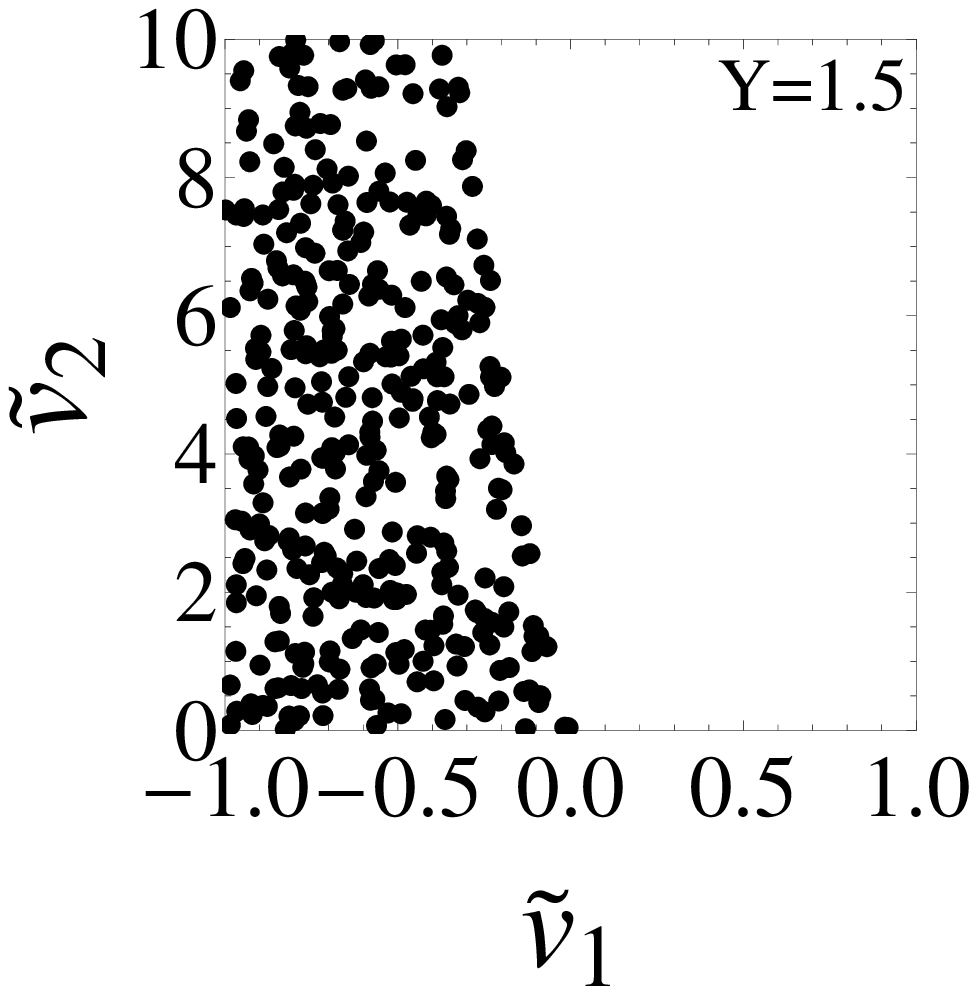,height=3.52cm,width=4.4cm,angle=0}
}
\centerline{\psfig{file=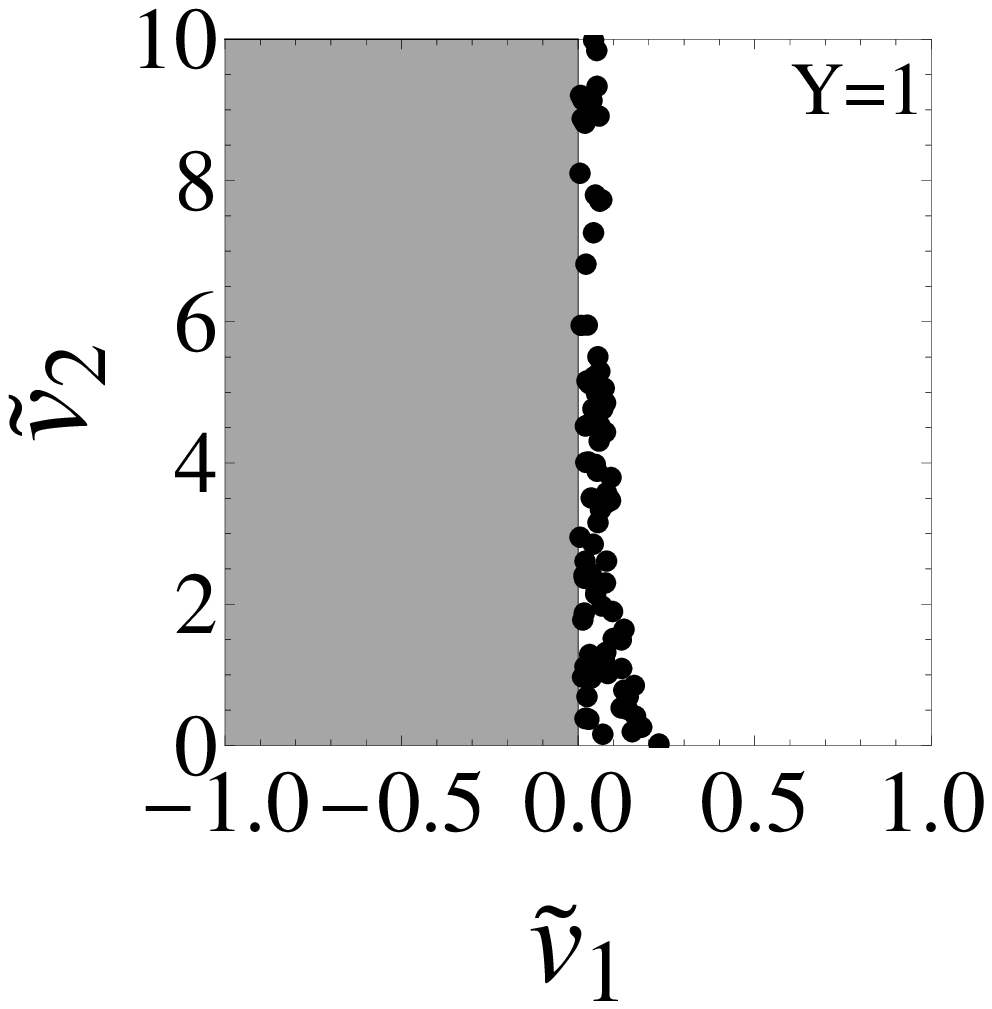,height=3.52cm,width=4.4cm,angle=0}
\psfig{file=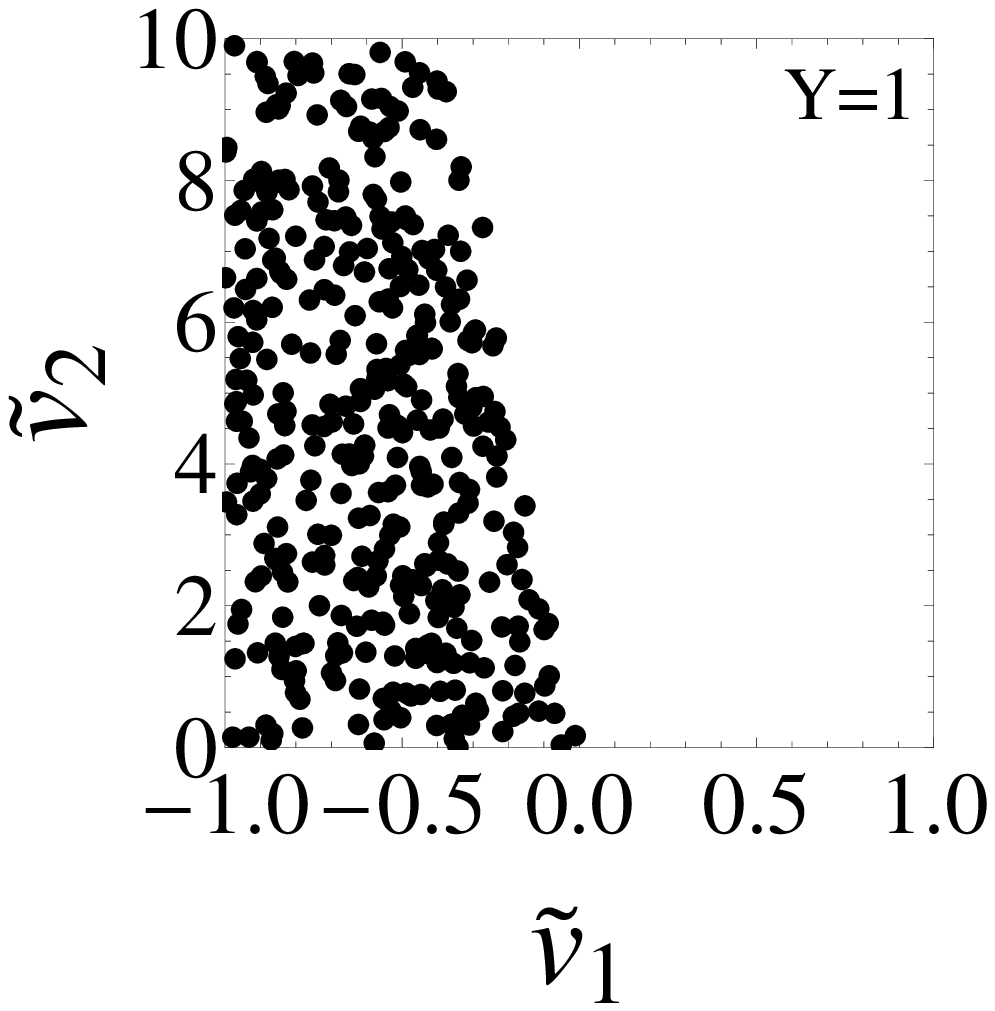,height=3.52cm,width=4.4cm,angle=0}
}
\centerline{\psfig{file=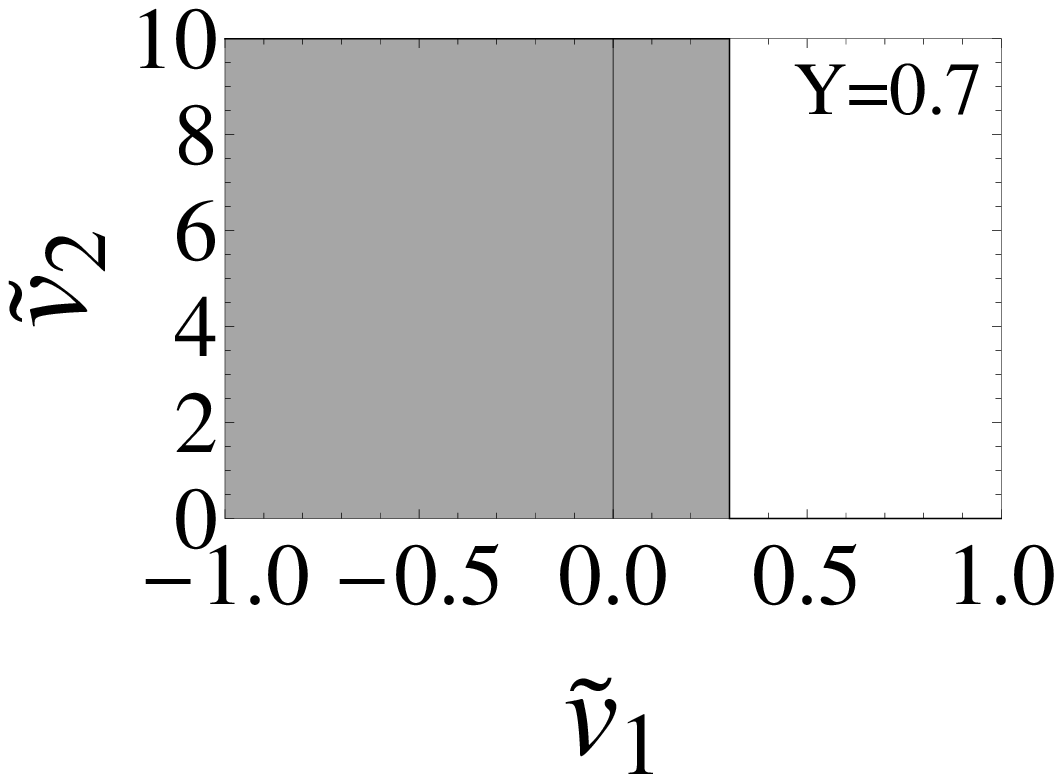,height=3.52cm,width=4.4cm,angle=0}
\psfig{file=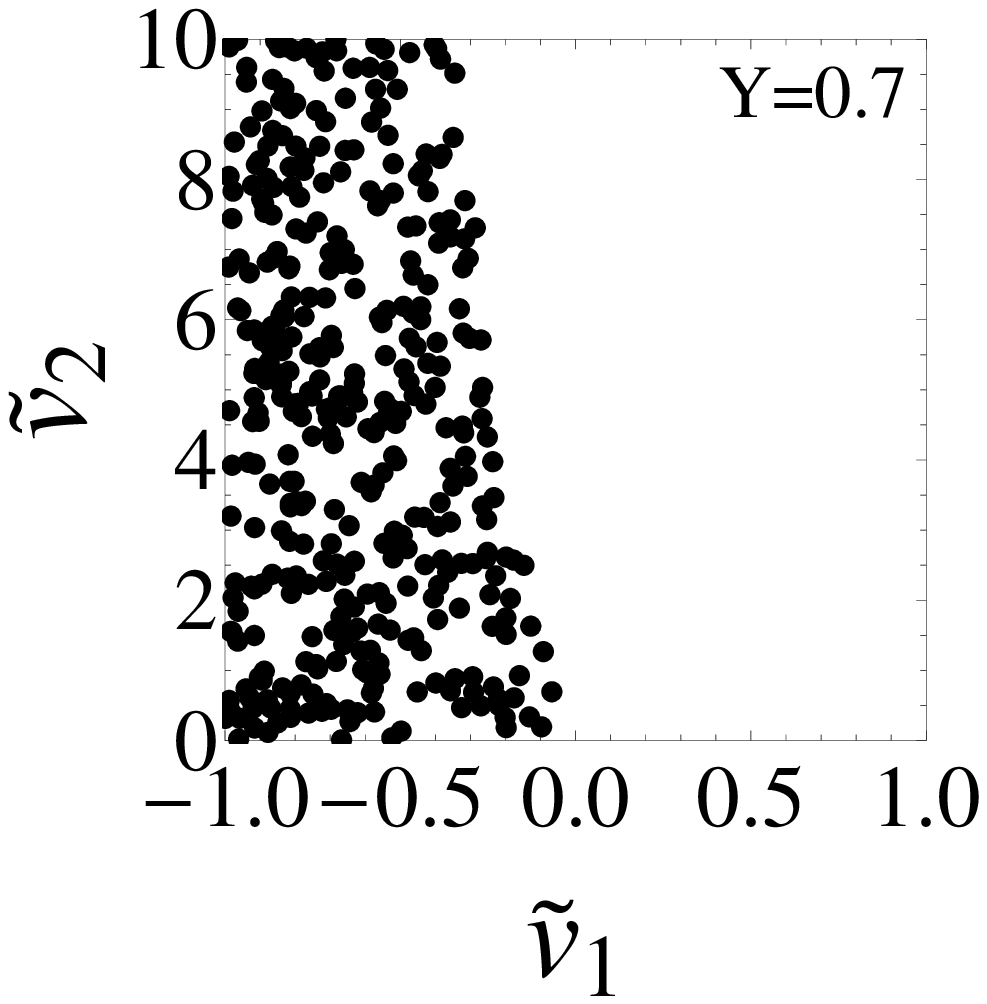,height=3.52cm,width=4.4cm,angle=0}
}
\centerline{\psfig{file=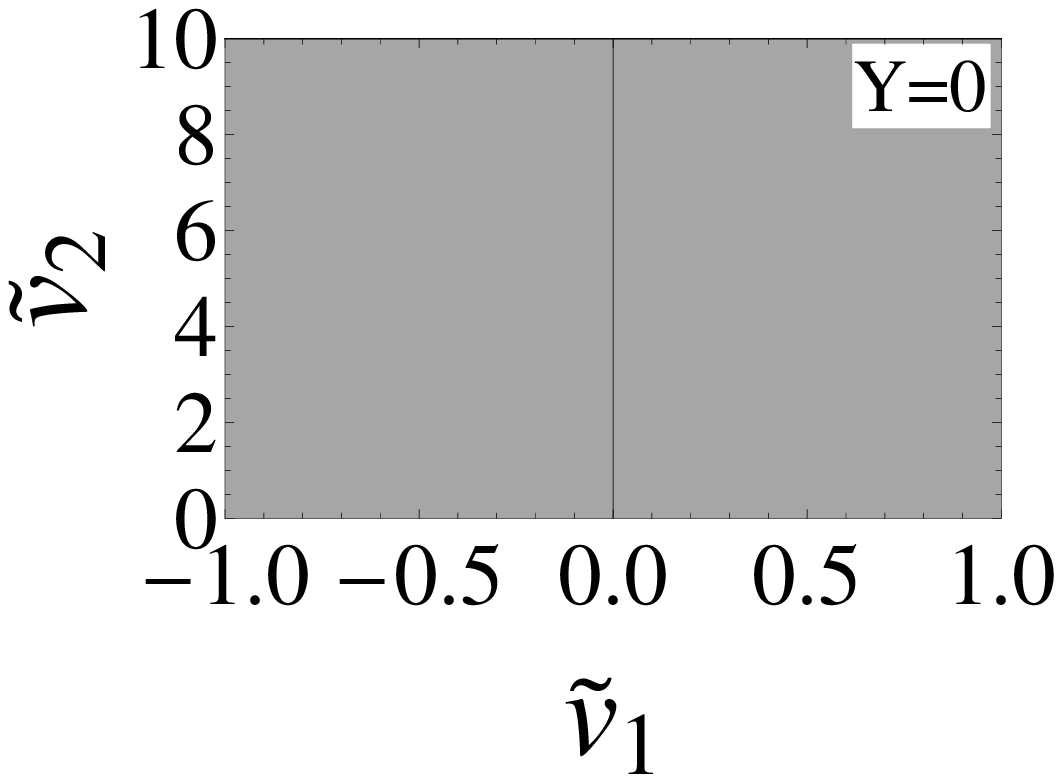,height=3.52cm,width=4.4cm,angle=0}
\psfig{file=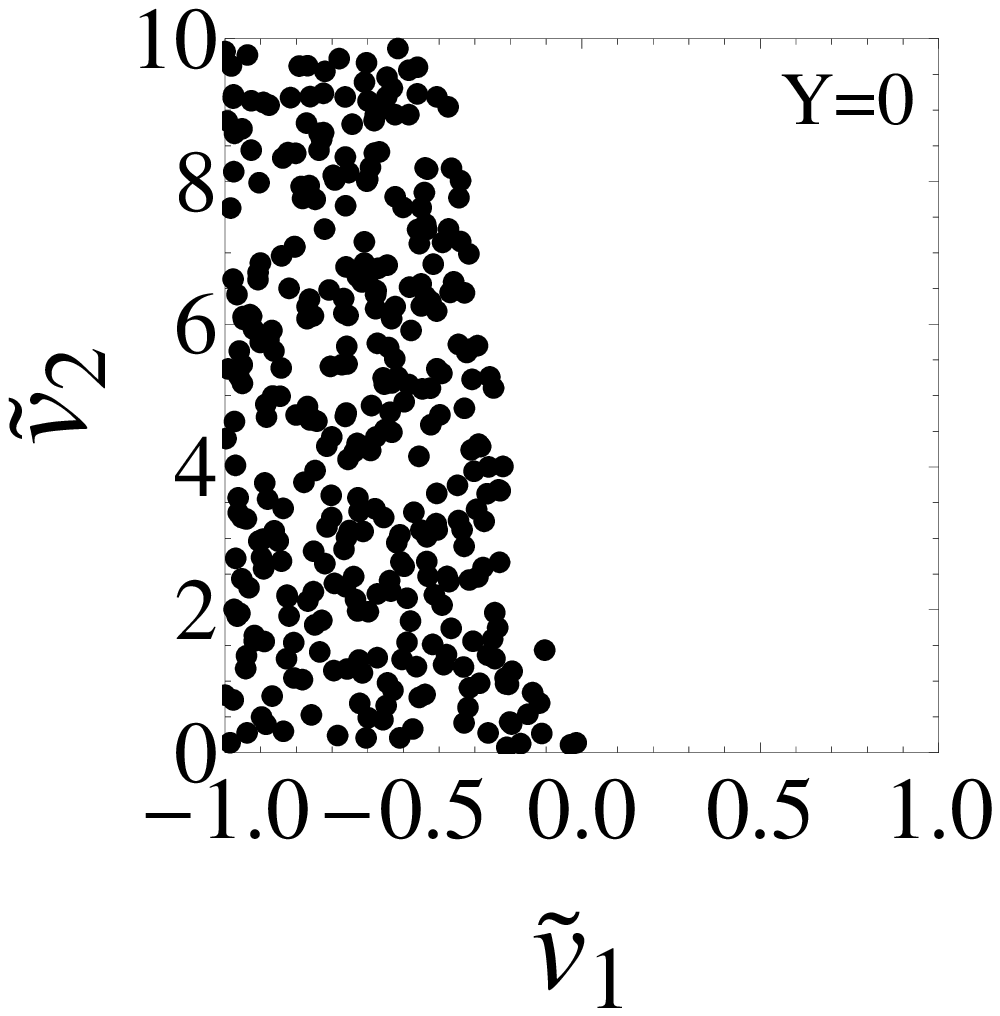,height=3.52cm,width=4.4cm,angle=0}
}
\caption{\label{dimywh} Phase diagrams in the plane $(\t{v}_1(\Lambda),\t{v}_2(\Lambda))$ for various given values $Y$ for the ghost (to the left) and the ordinary (to the right) $O(2)$ models. The empty regions correspond to the symmetric phase  I, the dotted and shadowed regions correspond to regions IIA and IIB of phase II, respectively.
}
\end{figure}

The WH RG flow is mapped numerically for keeping the dimensionful coupling $Y$ at various given constant values. The phases in the parameter plane 
 $\bigl(\t{v}_1(\Lambda),\t{v}_2(\Lambda)\bigr)$ of the bare dimensionless couplings can be distinguished by
considering the global RG flow started from the various points of that plane both for the ghost and ordinary $O(2)$ models.
In the parameter space the symmetric phase of the model, called here phase I, corresponds to the  points such that the RG trajectories started from them can be followed by the WH RG equation \eq{WHO2} down to the IR scale $k\to 0$, i.e., the
inverse propagator  $G^{-1}(k)\equiv s_-(k)|_{\Phi=0}=Zk^2+Yk^4+v_1(k)$ remains positive along those RG trajectories. As opposed to that, phase II is characterized by RG trajectories along which the inverse propagator either develops a zero at some finite scale $k_c$, $G^{-1}(k_c)=0$, at which the right-hand side of Eq. \eq{WHO2} becomes singular, or it is negative already at the UV scale, $G^{-1}(\Lambda)<0$. Therefore the region corresponding to phase II may consists of region IIA with
$G^{-1}(k)>0$ for $\Lambda\ge k >k_c>0$ 
 and region IIB with $G^{-1}(\Lambda)=Z\Lambda^2+Y\Lambda^4 +v_1(\Lambda)<0$, i.e., $-Z-Y>v_1(\Lambda)$ for $\Lambda=1$. Restricting ourselves to the bare parameter values satisfying the inequalities
$|v_1(\Lambda)|< \Lambda^2=1$ and $Y>0$, region IIB never occurs for the ordinary $O(2)$ model, but it occurs for the ghost model for $v_1(\Lambda)\le 1-Y$
when $Y<2$.

 Our first task is to find numerically the regions corresponding to phases I and  IIA in the parameter plane. The points of region IIA can be identified by
solving the  WH RG equation \eq{WHO2} and detecting that the inverse propagator vanishes at a finite scale $k_c$. For local potentials given in \eq{polpot} Eq. \eq{WHO2} reduces to the coupled set of ordinary first-order
 differential equations which has the form \eq{whv1v2} with $a=3$ and $b=5$ for $M=2$. 
In order to find region IIA  numerically, 1000 random starting points of the RG trajectories have been been generated in the parameter region
$\bigl( \t{v}_1(\Lambda), \t{v}_2(\Lambda)\bigr)\in [ -1,1]\otimes  [0, 10]$.
The phase diagrams are shown in Fig.   \ref{dimywh} for various values of
the higher-derivative coupling $Y$;
the empty, dotted, and shadowed regions
correspond to phase I, region IIA, and region IIB, respectively.
 
Numerics has revealed that phase II of the ghost model is bounded by $\t{v}_1(\Lambda)\le \t{v}_u(Y, \t{v}_2(\Lambda))$
while it is unbounded in the direction of $\t{v}_2(\Lambda)$ in the plane 
 $\bigl(\t{v}_1(\Lambda),\t{v}_2(\Lambda)\bigr)$.  For $Y\ge 1$ the phase boundary at $\t{v}_u>1-Y$  depends on the couplings $Y$ and $ \t{v}_2(\Lambda)$, so that region IIA also occurs with $1-Y< \t{v}_1(\Lambda)\le \t{v}_u$, while for $Y<1$ only region IIB occurs and $\t{v}_u=1-Y$. Thus the symmetric phase I
lies as a rule at larger values of  $\t{v}_1(\Lambda)$ in the parameter plane and for $Y\to  0$ it practically disappears. For the ordinary $O(2)$ model  phase II  contains only region IIA. The phase diagrams of the ordinary and ghost models are compared in Fig.  \ref{dimywh}. The phase boundary $\t{v}_u\bigl(Y,\t{v}_2(\Lambda)\bigr)$  depends approximately linearly on $\t{v}_2(\Lambda)$ as
$\t{v}_u\approx -c(Y) \t{v}_2(\Lambda)$ where $c(Y)$ is monotonically increasing with increasing value of the higher-derivative coupling $Y$. 
 Therefore the phase boundary is at $\t{v}_u\approx 0$ for  $\t{v}_2(\Lambda)\ll 1$ for all values $0\le Y\le 2$.

\subsubsection{IR scaling in phase I}
\begin{figure}[htb]
\centerline{\psfig{file=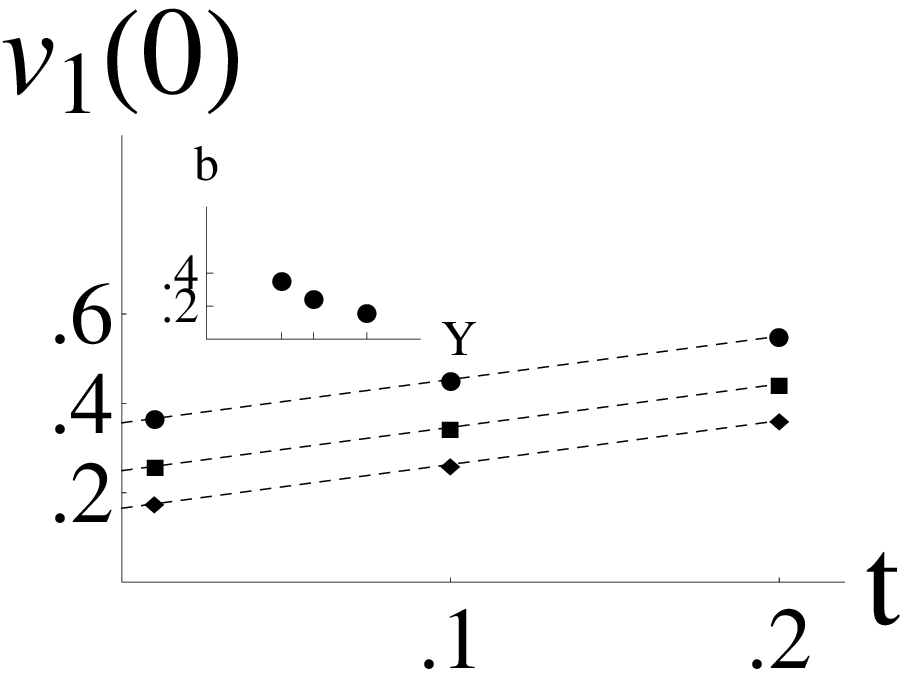,height=3.52cm,width=4.04cm,angle=0}\psfig{file=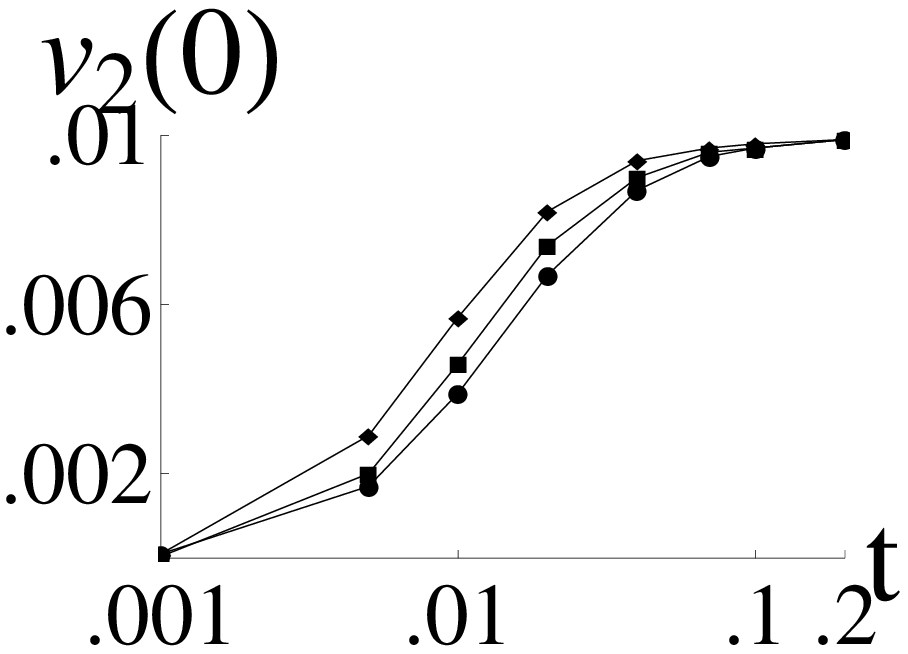,height=3.52cm,width=4.04cm,angle=0}
}
\caption{\label{fig:gh_ph1} The parameters $v_1(0)=at+b$ (to the left) and $v_2(0)$ (to the right)
vs. the `distance' $t=\t{v}_1(\Lambda)-\t{v}_u$ from the phase boundary for $v_2(\Lambda)=0.01$. The dots, boxes, and  rombs correspond to  $Y=0.7,~1.0,~1.5$, respectively, the lines are for guiding the eyes.
  The dependence of the coefficient $b$ on the higher-derivative coupling $Y$ is shown in the inset. 
}
\end{figure}

For phase I
 the IR scaling laws have been determined by solving the WH RG Eq. \eq{WHO2}.
The IR limits have been compared on
 RG trajectories started at various given `distances'  $t=\t{v}_1(\Lambda)-\t{v}_u$ from the phase boundary for $\t{v}_2(\Lambda)=0.01,~0.1$ and all investigated values of $Y$.
It has been found that the dimensionful couplings $v_n(k)$  of the local potential tend to constant nonvanishing values in the IR limit $k\to 0$ for both the 
ghost and the ordinary $O(2)$ models. 
The effective potential in phase I is  convex and paraboloid like for both the ghost and the ordinary models and  sensitive to the choice of the bare  potential. For the ghost model the linear relation 
\bea
  v_1(0)&=& a t +b(Y)
\eea
 has been established where the slope $a  $  is independent of $Y$, whereas
 the mass squared $b(Y)$ at the phase boundary for $t\to 0$ monotonically 
decreases with the increasing higher-derivative coupling  $Y$ (see Fig. \ref{fig:gh_ph1}).  The coupling $v_2(0)$  decreases with decreasing coupling $Y$ for given $t$, i.e., bare coupling $\t{v}_1(\Lambda)$ and approaching the phase boundary (i.e., for $t\to 0$) it tends to zero independently of the value of the coupling $Y$. For the ordinary counterpart of the  model the effective potential seems  to be insensitive to the value of the higher-derivative coupling in the range $0\le Y <2$, but keeps its sensitivity to the parameters of the bare potential.

\subsubsection{IR scaling laws in phase II}

The RG trajectories belonging to region IIA can be followed by the WH RG equation \eq{WHO2} from the UV scale $\Lambda$ to the scale $k_c$ of the singularity  and 
  the scaling of the couplings in the deep IR region $k<k_c$ should be
  obtained by TLR which has been started from the initial potential obtained at the critical scale $k_c$ by the solution of Eq. \eq{WHO2}. In order to find the RG trajectories belonging to region IIB   TLR should be started at the UV scale. In both cases the ansatz \eq{polpot} with the truncation $M=10$ has been used for the potential. In our numerical TLR procedure
 the scale $k$ has been decreased from either the critical one ($k_c$) for region IIA or from the UV scale $\Lambda$ for region IIB  by $3$ orders of magnitude with the  step size $\Delta k /k = 0.01$ during the numerical tree-level blocking. Generally $\sim 500$ iteration  steps have been numerically performed at each value of the constant background $\Phi$ for the minimization of the blocked potential $U_k(\rho, \Phi)$
with respect to the amplitude $\rho$ of the spinodal instability.
The TLR procedure  is quantitatively sensitive to the choice of the interval
$|\Phi| \le \b{\Phi}$ in which the minimization of the potential $U_k(\rho, \Phi)$
with respect to the amplitude $\rho$ of the spinodal instability and the least square fit
of the blocked potential at scale $k-\Delta k$ are performed. For `Mexican hat'
like potential $U_{k_c}(\Phi)$ for region IIA or $U_\Lambda(\Phi)$ for region IIB the choice $\b{\Phi}\approx 1.5 \Phi_m$ has been made where $\pm\Phi_m$ are the
 positions of the local minima of the potential with $\Phi_m=
 \sqrt{ -2v_1(k_c)/v_2(k_c)} $ or $\Phi_m=
 \sqrt{ -2v_1(\Lambda)/v_2(\Lambda)} $, respectively.
For convex potentials with $v_1(k_c)>0$ for region IIA or $v_1(\Lambda)>0$ for
region IIB the choice $\b{\Phi}\gtrsim 30$ has been made. It has been observed
numerically that the blocked potential does not acquire tree-level corrections outside of the interval $|\Phi|\le \Phi_c$  with $\Phi_c$ given by Eq. \eq{phico2}, but the choice of the larger interval makes the minimization and fitting
numerically stable.

\begin{figure}[thb]
\centerline{\psfig{file=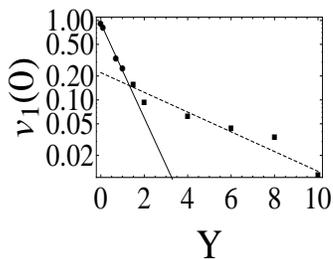,height=3.52cm,width=4.4cm,angle=0}
}
\caption{\label{fig:log_g2againsty} The dependence of mean values
 $\overline{v_1(0)}$ on the higher-derivative coupling $Y$ in phase II.
The lines are only to guide the eyes. 
}
\end{figure}

For each given value of the higher-derivative coupling $Y$ and both values of the bare coupling $\t{v}_2(\Lambda)=0.01$ and $0.1$ we have determined the RG trajectories for 3 to 5 bare values of $\t{v}_1(\Lambda)$ distributed uniformly in the interval $-1<\t{v}_1(\Lambda)<\t{v}_u$.
It was found that the couplings of the dimensionful blocked potential tend to constant values in the IR limit. Moreover, it has been observed that
for any given value of the higher-derivative coupling $Y$
 the effective potential  is universal  in the sense that it does not depend on at which point $(v_1(\Lambda),~v_2(\Lambda))$ the RG trajectories have been started.
  Therefore we have determined the mean values $\overline{v_1(0)}$ and $\overline{v_2(0)}$ of the couplings $v_1(0)$ and $v_2(0)$ with their variances via averaging them over all evaluated RG trajectories belonging to a given value of the coupling  $Y$. It turned out that
the mean value $\overline{v_1(0)}$ decreases strictly monotonically with increasing values of the higher-derivative coupling $Y$ as shown in Fig. \ref{fig:log_g2againsty}. The mean values $\overline{v_2(0)}$ take randomly positive and negative small values with variances comparable with their magnitudes when the coupling $Y$ is altered. Thus we concluded that the quartic coupling of the effective potential vanishes, an averaging over all considered values of the coupling $Y$ yields
$\la  \overline{v_2(0)}\ra =0.004\pm  0.01$.
Therefore the dimensionful effective potential is an upsided paraboloid with its minimum at $\Phi=0$ in phase II. Moreover, the nonrenormalizable, UV irrelevant coupling $Y$  turns out to be IR relevant in phase II.

For each RG trajectory we have also evaluated the ratio $r$ characterizing how large part of the sum of the negative terms is cancelled totally or partially by the positive higher derivative term  in the inverse propagator $G^{-1}(k_s)$,
\bea
   r&=& \Biggl\{ 
\begin{array}{ccc}
\frac{ Yk_s^4}{Yk_s^4+v_1(k_s)}, &  {\mbox{if}} &  v_1(k_s)\ge 0
\cr
  1              ,  &   {\mbox{if}} &  v_1(k_s) < 0
\end{array},
\eea
where $k_s=k_c$ and $k_s=\Lambda$ for regions IIA and IIB, respectively.
  In this manner the ratio $r$ characterizes how significant is the  role played by the ghost condensation  in this cancellation either at  scale $k_s$ where
TLR should be started.
In the cases with $v_1(k_s)<0$ the ghost condensation is the only mechanism that can 
be responsible for the above mentioned cancellation. Numerics has shown that
the value of the ratio is at $r\approx 1$ in most of the parameter region belonging to phase II, but it generally decreases suddenly when $\t{v}_1(\Lambda)$
 approaches the phase boundary at $\t{v}_u$. The IR couplings in the effective potential seem, however, to be insensitive to the value of $r$, i.e., 
 to the importance of the ghost condensation at the  scale $k_s$.
As argued for below, rather the role
of the ghost condensation during the {\em global} RG flow during TLR makes its imprint on the value of $v_1(0)$ via its dependence on the coupling $Y$.

\begin{figure}[htb]
\centerline{\psfig{file=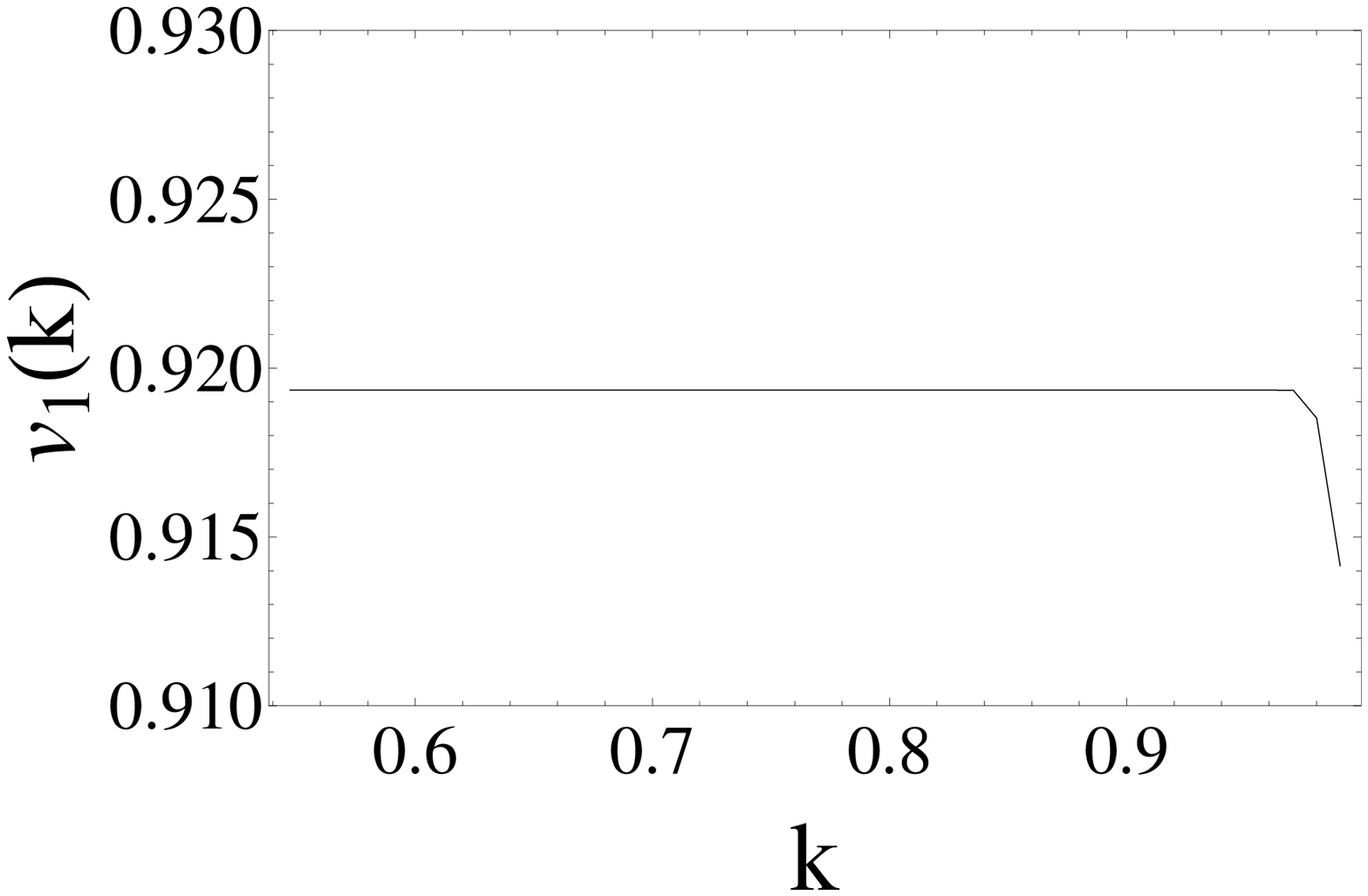,height=3.52cm,width=4.4cm,angle=0}\psfig{file=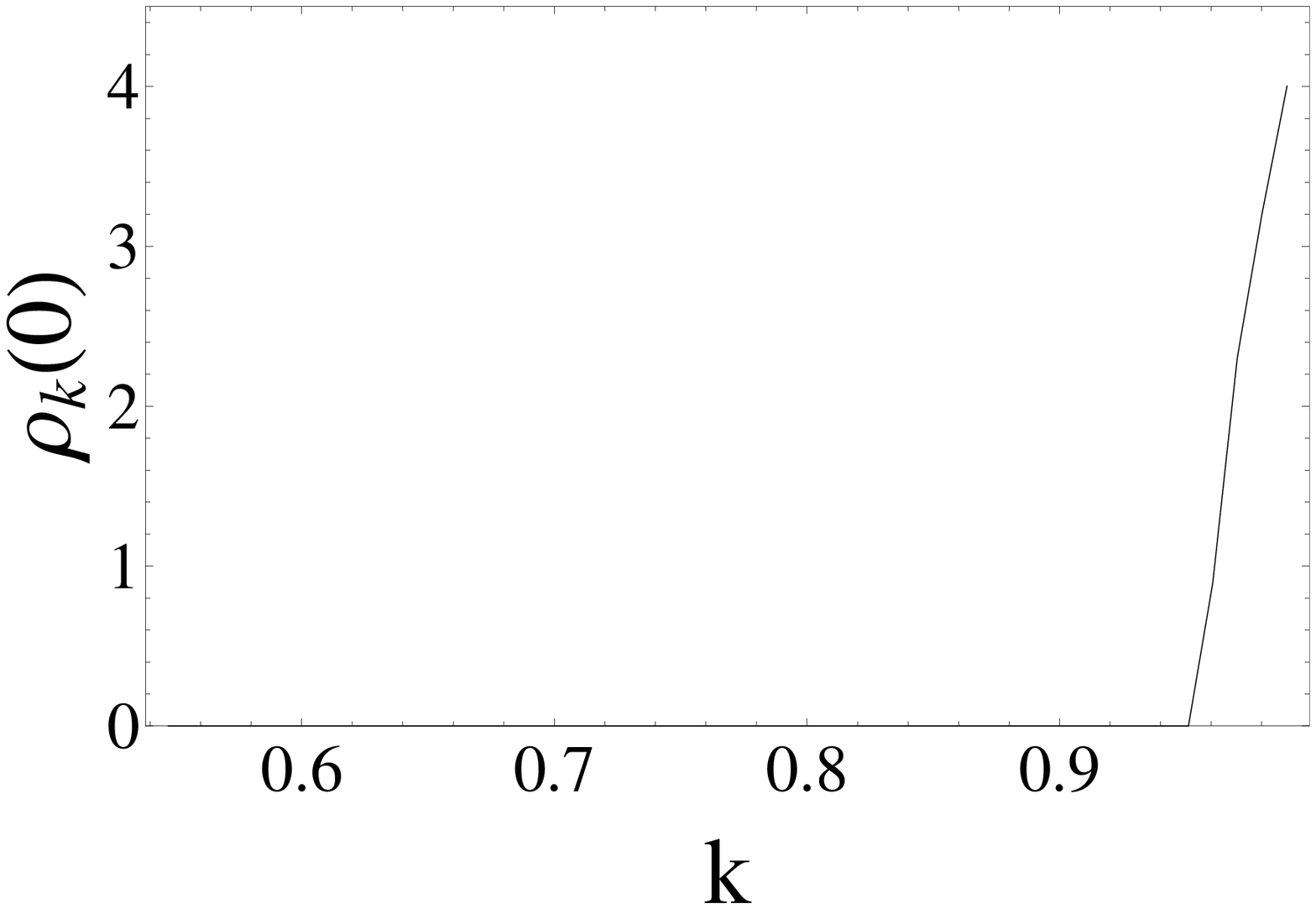,height=3.52cm,width=4.4cm,angle=0}}
\centerline{\psfig{file=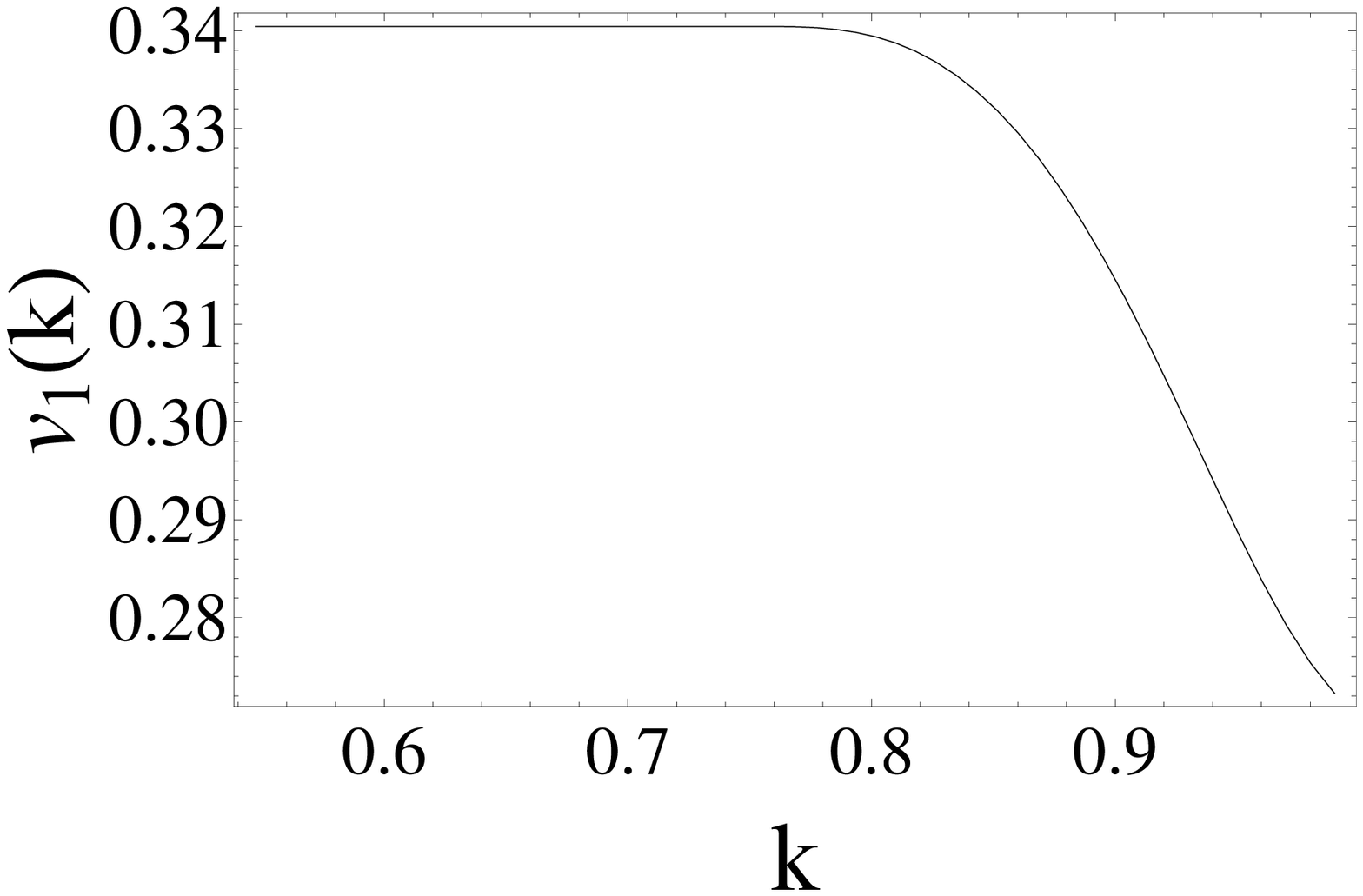,height=3.52cm,width=4.4cm,angle=0}\psfig{file=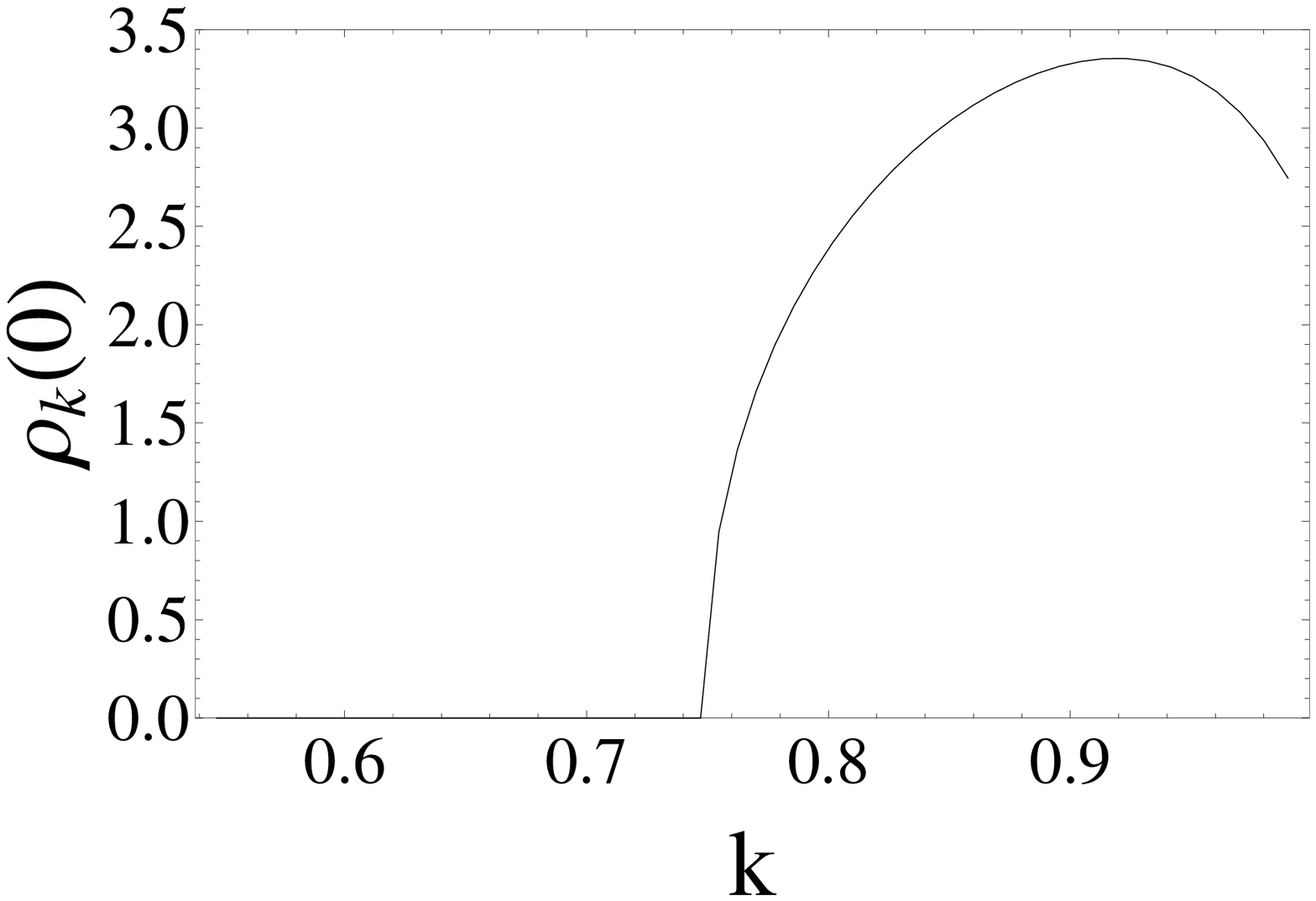,height=3.52cm,width=4.4cm,angle=0}}
\centerline{\psfig{file=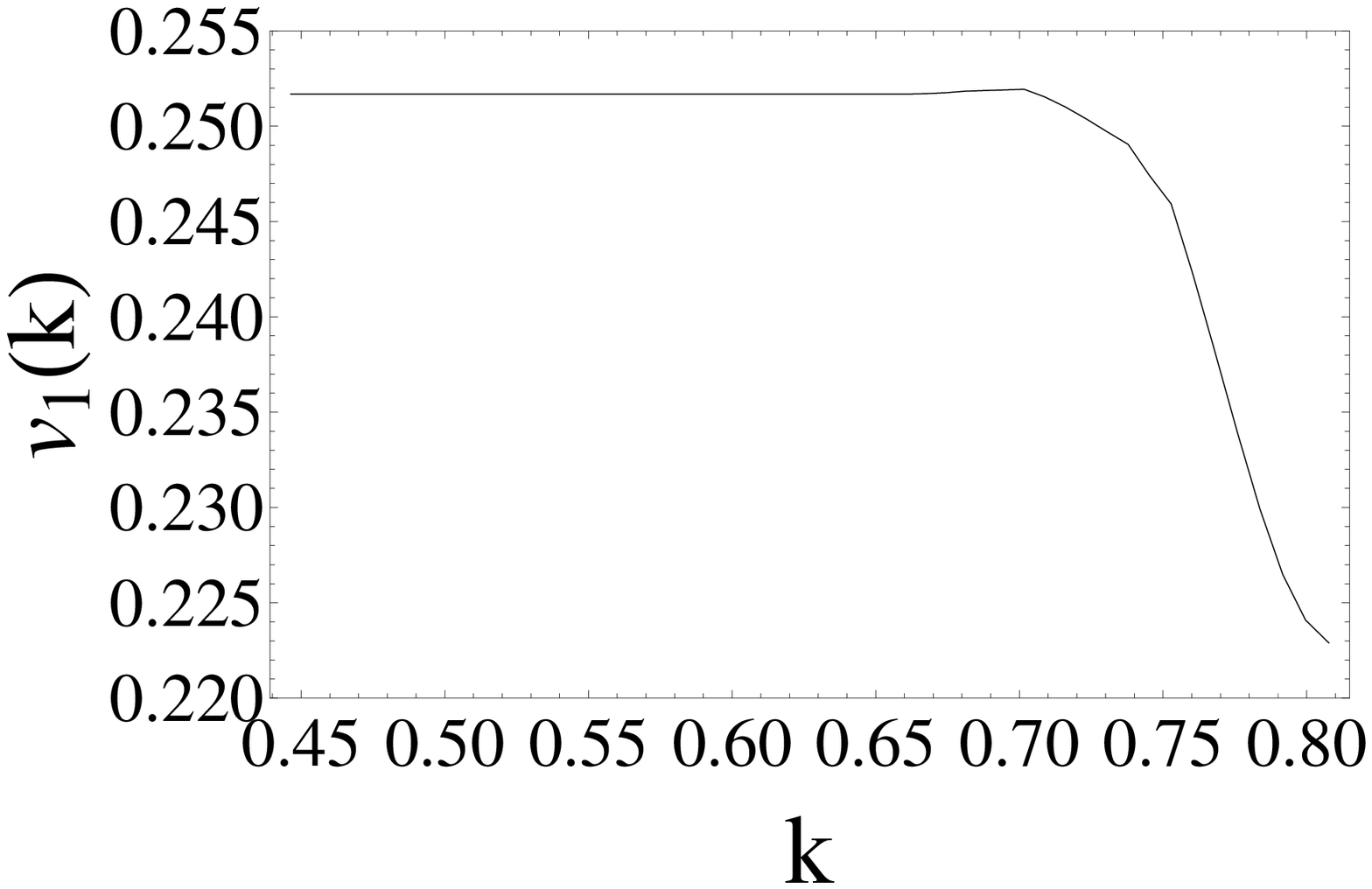,height=3.52cm,width=4.4cm,angle=0}\psfig{file=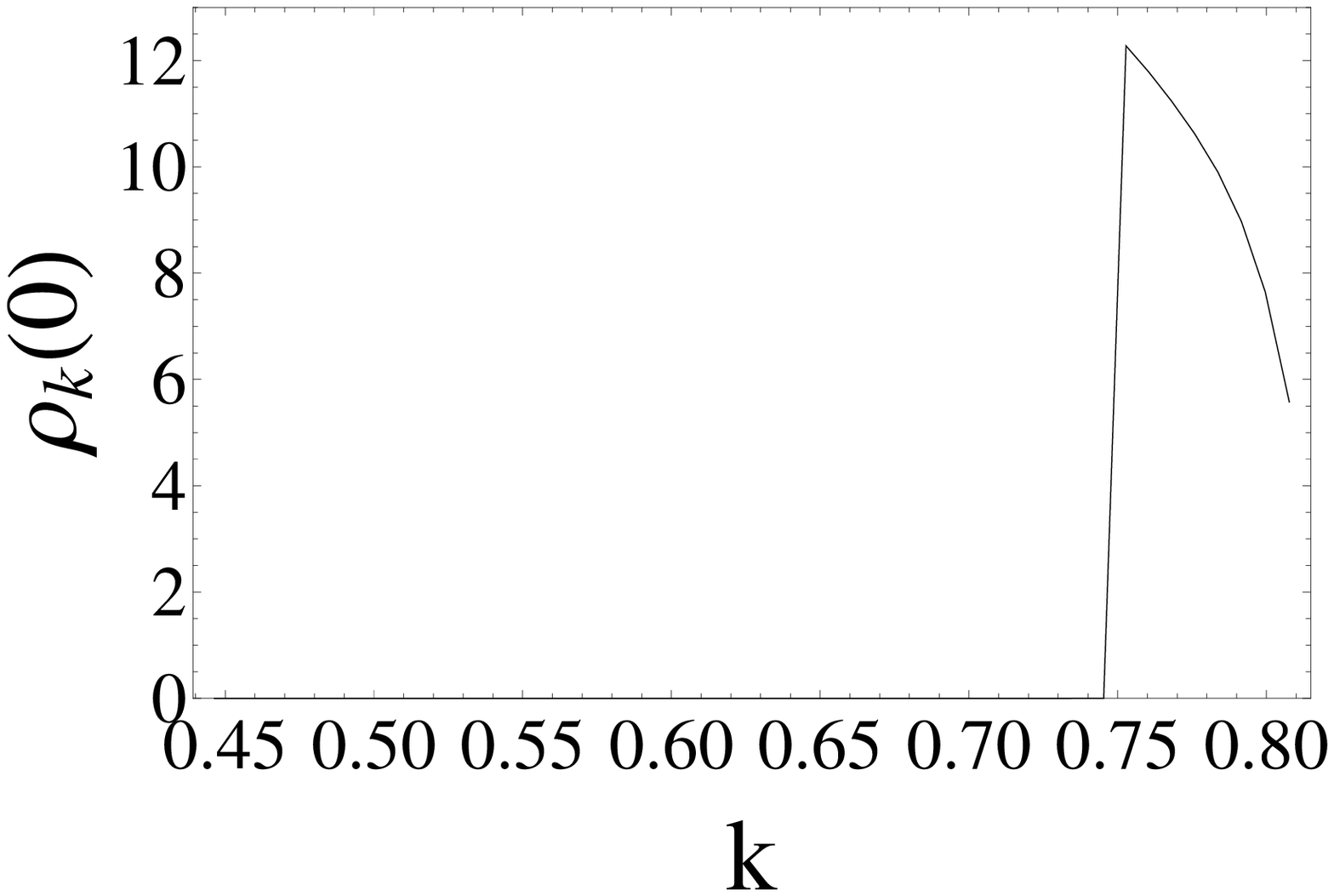,height=3.52cm,width=4.4cm,angle=0}}
\centerline{\psfig{file=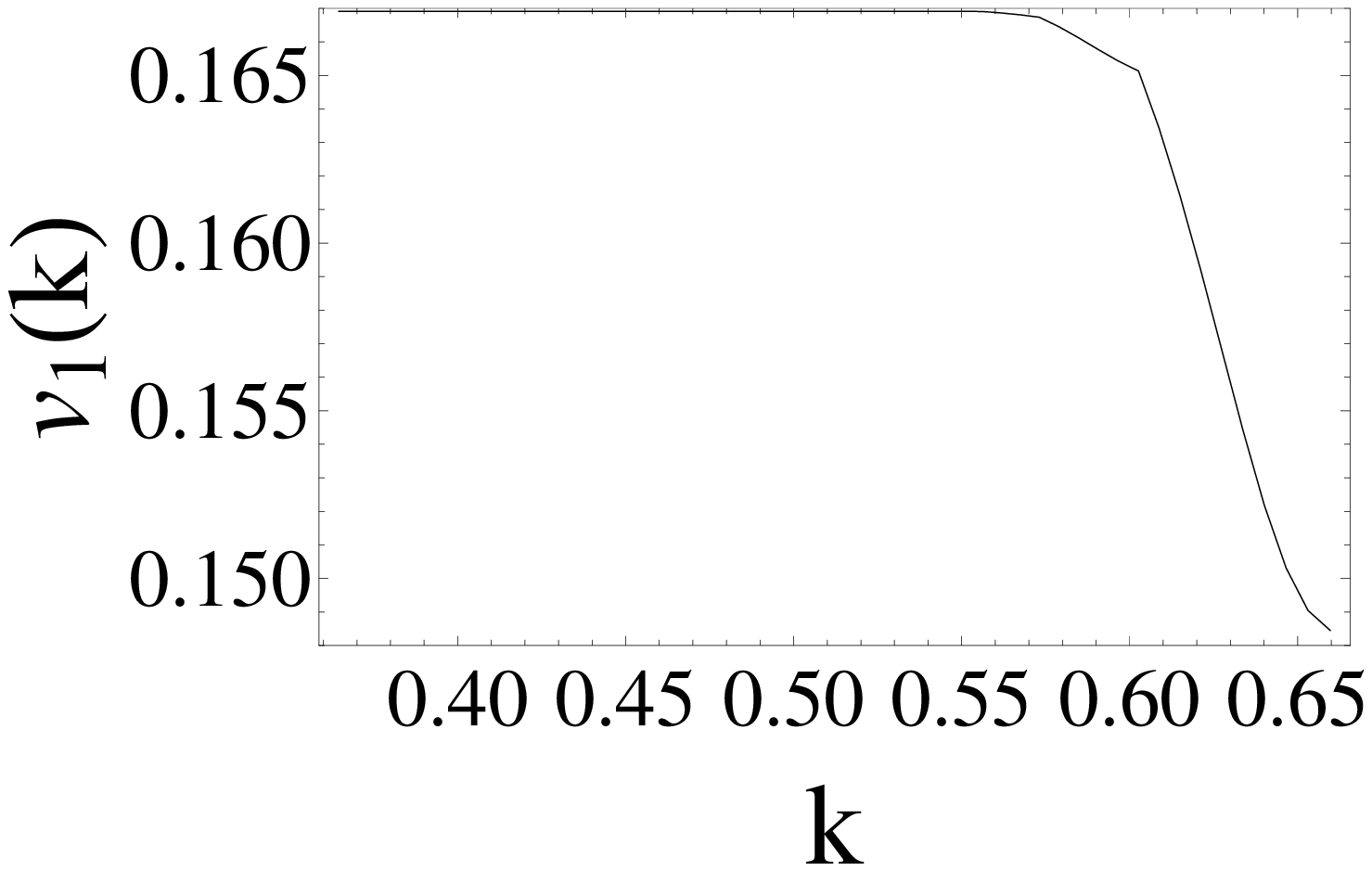,height=3.52cm,width=4.4cm,angle=0}\psfig{file=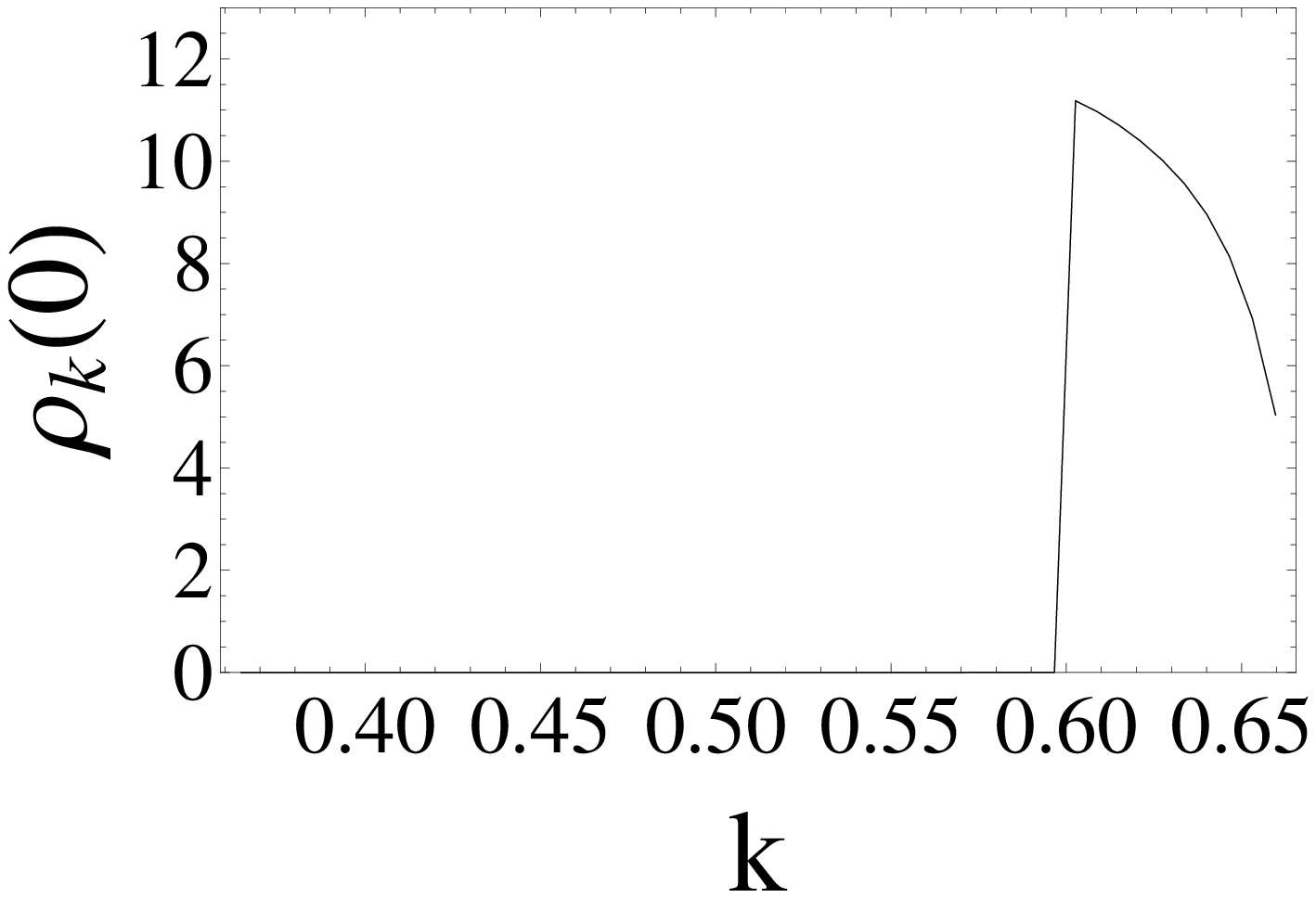,height=3.52cm,width=4.4cm,angle=0}}
\caption{\label{fig:irsc_ph2_v1rho}Thr TLR flow of $v_1(k)$ and the corresponding $\rho_k(\Phi=0)$ 
for  $t=\t{v}_2(\Lambda)=0.1$ and $Y=0,~0.7,~1.0$, and $1.5$ (from the top to the bottom) in phase II of the ghost $O(2)$ model.
}
\end{figure}

 The numerical TLR procedure has shown that although the amplitude $\rho_k$ of the spinodal instability suddenly acquires a large value
 just below  either the scale  $\Lambda$ for region IIB
(cases $Y=0$ and $0.7$ in  Fig. \ref{fig:irsc_ph2_v1rho}) or
the scale $k_c$ for region IIA (cases $Y=1.0$ and $1.5$  in  Fig. \ref{fig:irsc_ph2_v1rho}), but after relatively few  $(\sim 30)$ blocking steps it is
rapidly washed away  and
does not survive the IR limit. The vanishing of $\rho_k(0)$ is accompanied by
the saturation of the value of $v_1(k)$ at its IR limiting value $v_1(0)$.
 In Fig.  \ref{fig:irsc_ph2_v1rho} the plots
belong to RG trajectories characterized by the ratio $r\approx 1$. 
 This indicates that  basically the ghost condensation should be responsible
for the occurence of the finite amplitude $\rho_k(\Phi)$ of the spinodal instability when TLR is started. However, as numerics shows, the RG flow of the local potential
starts to dominate the IR scaling  after a relatively small decrement of the scale $k$.  Nevertheless, this would-be condensate
 seems to left behind its footprint on the curvature of the effective potential
 through the dependence of the mass parameter $v_1(0)$ on the higher-derivative
 coupling $Y$. 
As seen  in Fig. \ref{fig:log_g2againsty}, the exponential dependence of $\overline{v_1(0)}$ on  $Y$ changes its slope at around $Y\approx 1$.
Fig.  \ref{fig:irsc_ph2_v1rho} seems to support the conjecture that
the ghost condensation plays the most significant role during the global RG flow when the higher-derivative coupling is at around $Y\approx \ord{\Lambda^{-2}=1}$. Namely, Fig.  \ref{fig:irsc_ph2_v1rho} shows that
 the width of the $k$-interval in which $\rho_k(\Phi=0)$ is nonvanishing increases for $Y$ increasing from 0 towards 1, but it remains unaltered for $Y\gtrsim 1$. 
This may be connected with the following
circumstances. The kinetic piece $\Omega(k^2)$  of the inverse propagator is an upsided parabola with zeros at $k^2=0$ and $k^2=1/Y$ and the minimum at $k^2=1/(2Y)$. If $Y\gg \Lambda^{-2}=1$ then the modes which can give negative contributions to the action by ghost condensation represent a small amount of the modes below the UV cutoff $\Lambda=1$. In the extreme limit $Y\to \infty$ these modes
 are restricted to an interval of vanishing size at zero momentum, and the spinodal instability is governed by the potential. In the other extreme with $Y\ll \Lambda^{-2}=1$ all modes below the UV cutoff are available for ghost condensation, but for $1/(2Y)\gg 1$ all they may give rather small negative contribution to the action and in the limit $Y\to 0$ this contribution becomes negligible. By  this naive argumentation one concludes that the ghost condensation may only play a significant
role in forming the IR value $v_1(0)$ for $Y\approx \ord{\Lambda^{-2}=1}$. 

It has also been established numerically that the range $\Phi_c(k)$ of the homogeneous background field in which spinodal instability occurs opens up gradually when the scale $k$ decreases from $k_s$, its width reaches a maximum and then
suddenly decreases to zero at some finite scale $k_0$, where also the amplitude $\rho_k$ vanishes and the couplings $v_1(k)$ and $v_2(k)$ reach their IR values. No more TLR corrections appear below that scale $k_0$. This behaviour is rather different of that shown up by the ordinary $O(2)$ symmetric model in its broken symmetric phase.

 The ghost condensate occurring  at the scale $k_s$ breaks internal $O(2)$ symmetry as well as Euclidean rotational symmetry in the 3-dimensional space and translational symmetry in the $x_1$ direction in the Euclidean space. These symmetries are, however, restored in the IR limit. Thus we have to conclude that even  phase II of the ghost model is a symmetric one. The distinction between phases I and II can only be done by considering the global RG flow: the effective potential exhibits no sensitivity to the couplings of the bare potential in phase II, as opposed to phase I, where such a sensitivity is essential.

It has also been checked numerically that for phase II of the ordinary $O(2)$ model with nonvanishing higher-derivative coupling $Y$ the TLR reproduces the Maxwell-cut for the dimensionful effective potential, as expected.

\subsubsection{Correlation length}
\begin{figure}[thb]
\centerline{\psfig{file=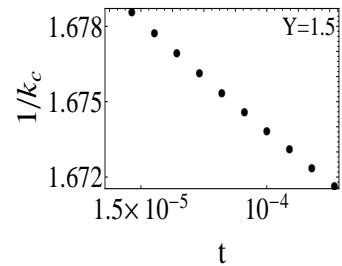,height=3.52cm,width=4.4cm,angle=0}
}
\caption{\label{fig:critical_exp} Scaling of the correlation length $\xi\sim 1/k_c$ 
with the reduced temperature $t=\t{v}_u-\t{v}_{1}(\Lambda)$ (on a lin-lin plot) at the boundary of
 phases I and II of the ghost $O(2)$ model for $Y=1.5$ and $\t{v}_2(\Lambda)=0.1$.
}
\end{figure}

\begin{figure}[thb]
\centerline{\psfig{file=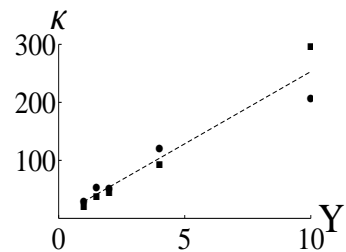,height=3.52cm,width=4.4cm,angle=0}
}
\caption{\label{fig:alpha_ph2} The coefficient $\kappa$ in  expression
\eq{corpowlaw} of the correlation length vs.  the higher-derivative coupling $Y$. The points correspond to various RG trajectories, the line is for guiding the eyes. 
}
\end{figure}

\begin{figure}[thb]
\centerline{\psfig{file=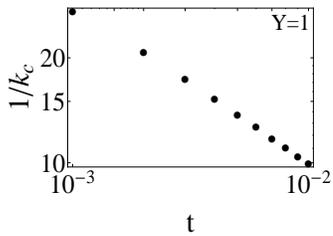,height=3.52cm,width=4.4cm,angle=0}
}
\caption{\label{fig:ord_crit_exp}  Typical scaling of the correlation length $\xi\sim 1/k_c$ 
with the reduced temperature $t=\t{v}_u-\t{v}_{1}(\Lambda)$ (on a log-log plot)  at the boundary of
 phases I and II of the ordinary $O(2)$ model for $Y=1.0$ and $\t{v}_2(\Lambda)=0.1$. The critical exponent is $\nu\approx 0.45\pm 0.02$ in the range $Y\in [0, 1.5]$.
}
\end{figure}

Finally, let us determine
the behaviour of the correlation length $\xi\sim 1/k_c$ approaching the boundary of phases I and II from the side of phase II for the ghost model. This is only possible for the RG trajectories belonging to region IIA, when the singularity scale
$k_c$ can be detected by solving the WH RG equation \eq{WHO2}. 
Fig.  \ref{dimywh} makes it plausible that for given $\t{v}_2(\Lambda)$ the `distance' $\t{v}_u-\t{v}_{1}(\Lambda)$  measures how far an RG trajectory belonging to phase II runs from the boundary  of the phases I and II.  Therefore, one can identify $\t{v}_u- \t{v}_{1}(\Lambda)$  with the reduced temperature $t$ up to a constant factor. In order to determine the dependence of the correlation length $\xi$ on the difference  $\t{v}_u- \t{v}_{1}(\Lambda)$,
 we have solved
 the WH RG equations with various initial conditions  $\t{v}_2(\Lambda)=0.01,~0.1$ and  $\t{v}_{1i}(\Lambda)=[1-(i/100)]\t{v}_u $ $(i=1,2,\ldots)$ for each values  $Y=1.0,~1.5,~2.0,~4.0,~10.0$. 
It has been established that the correlation length increases linearly with
decreasing reduced temperature,
\bea\label{corpowlaw}
   \xi\sim 1/k_c&=& \xi_0 -\kappa [\t{v}_u-\t{v}_{1}(\Lambda)],
\eea
for any fixed values the coupling $Y$, as shown in  Fig. \ref{fig:critical_exp}.
The coefficient $\kappa$ seems to rise nearly linearly with increasing higher-derivative coupling $Y$ (see Fig. \ref{fig:alpha_ph2}). 
Although the correlation length increases approaching the phase boundary from the side of phase II, but it remains finite, while it is infinite in the symmetric phase I.  This signals that the phase transition  of the ghost $O(2)$ model  is of first order, as opposed to the ordinary $O(2)$ model where the correlation length blows up according to the power law
$\xi\sim t^{-\nu}$  (see Fig. \ref{fig:ord_crit_exp}), indicating a continuous phase transition.

\section{Summary}\label{sec:sum}

In the present paper the phase structure and the IR behaviour of the $O(2)$ 
symmetric ghost scalar field model with the kinetic energy operator
$\Omega(-\Box)=-Z\Box+ Y\Box^2$ with $Z=-1$, $Y>0$
 has been investigated in 3-dimensional
Euclidean space in the framework of Wegner and Houghton's renormalization group 
(WH RG) scheme. A particular emphasis has been laid on tree-level renormalization (TLR) in order to obtain the deep IR scaling of the blocked potential. The opposite signs
of the wavefunction renormalization $Z$ and the higher-derivative coupling $Y$ 
enable  the system to lower the value of its action for inhomogeneous field configurations as compared to that for homogeneous ones. The corresponding nontrivial saddle-points of the  path integral have been looked for in sinusoidal  form 
of wavelength $2\pi/k$ in one spatial direction, where $k$ is the gliding sharp cutoff scale of the WH RG approach. The occurrence of such a periodic field configuration breaks -- among others -- the internal $O(2)$ symmetry spontaneously.
The WH RG approach has been applied by keeping the dimensionful higher-derivative coupling $Y$ constant.

Our numerical TLR procedure has been tested by successfully reproducing well-known results for the symmetry broken phase of ordinary ($Z=+1$, $Y=0$) one-component real scalar field model in 3-dimensional Euclidean space and for the molecular phase of (ordinary) sine-Gordon model in 2-dimensional  Euclidean space.

It has been established that the 3-dimensional $O(2)$ symmetric ghost scalar model has two phases. In phase I, i.e., in the symmetric phase no ghost condensation occurs; the couplings of the dimensionful blocked  potential tend to constant values in the IR limit $k\to 0$, their values depend  on the couplings of the bare potential and on the higher-derivative coupling $Y$. The dimensionful effective potential for phase I is of paraboloid-like shape. In phase II spinodal instability occurs along the RG trajectories at some finite scale $k_s$ which was shown to occur basicly due to ghost condensation. Numerics has revealed, however, that neither the amplitude of the condensate nor the interval of the homogeneous background field in which it occurs  survive the IR limit.
This means that the symmetries broken at intermediate scales are restored in the IR limit.
 Nevertheless, this would-be condensate makes a significant imprint in the effective potential which becomes insensitive to the couplings of the bare potential. The curvature of the effective potential decreases monotonically with increasing higher-derivative coupling $Y$, while the couplings of its  higher-order terms
 vanish, so that the effective potential of phase I with restored symmetry
is of paraboloid shape. The  identification of the scale of spinodal instability in phase II with the reciprocal of the correlation length has shown  that
 approaching the phase boundary the correlation length  increases to a {\em finite} value,  indicating that the ghost model exhibits a phase transition of the first order.

The phase structure obtained for the 3-dimensional ghost $O(2)$ model
has been compared to that of the ordinary ($Z=+1$, $Y>0$) $O(2)$ model. The 
latter has a symmetric phase and a symmetry broken one. In the latter the
 effective potential is insensitive to the bare potential and reproduces the Maxwell cut. The correlation length has been found to scale according to a continuous phase transition.

\section*{Acknowledgements}
S. Nagy acknowledges financial support from a J\'anos Bolyai Grant of the
Hungarian Academy of Sciences, the Hungarian National Research Fund OTKA
(K112233).

\appendix

\section{Tree-level renormalization of Euclidean one-component scalar field theory with polynomial potential}\label{treelerg}

Here we would like to remind the reader how TLR works in
one-component scalar field theory with ordinary kinetic term. More detailed discussion can be found in Ref. \cite{Ale1999}.
For scales $k<k_c$ spinodal instability occurs when the logarithm in the right-hand side of Eq.  \eq{WH} satisfies
 the inequality  
\bea\label{splusle0}
 Z+\t{v}_1(k_c)+\frac{3}{2}\t{v}_2(k_c)\t{\Phi}^2&\le &0.
\eea
Since the last term in the left-hand side of the inequality \eq{splusle0}
is never negative, the critical scale is given via the equation
\bea
 Z+\t{v}_1(k_c)&=&0.
\eea
Moreover, one can estimate the interval $\Phi\in [-\Phi_c(k),\Phi_c(k)]$ in which
instability occurs for scales $k<k_c$ from inequality \eq{splusle0} as
\bea\label{phicone}
&&\t{\Phi}_c(k)= \sqrt{-\frac{ 2\lbrack Z+\t{v}_1(k)\rbrack}{3\t{v}_2(k)}},~~
 \Phi_c(k)=\sqrt{k}\t{\Phi}_c(k) .
\eea
For scales $k<k_c$ and background fields $\Phi\in [-\Phi_c(k),\Phi_c(k)]$ one turns to the tree-level blocking relation \eq{blocktree} and inserting the ansatz \eq{polpot} into it, one obtains the recursion relation
\bea\label{blocktree2}
U_{k-\Delta k}(\Phi)&=&\min_{\{\rho\}}\biggl(U_k(\Phi)+Zk^2\rho^2\nn
&&+\sum_{n=1}^M \frac{\rho^{2n}}{(n!)^2}\partial_\Phi^{2n} U_k(\Phi)\biggr)
\eea 
for the blocked potential. For given scale $k$ with given couplings $v_n(k)$ and
for given homogeneous field $\Phi\in [-\Phi_c(k),\Phi_c(k)]$, one determines
 the value $\rho_k(\Phi)$ minimizing the right-hand side of Eq.  \eq{blocktree2}. Then one repeats this minimization for various $\Phi$ values and determines the corresponding $U_{k-\Delta k}(\Phi)$ values. Finally these discrete values of   $U_{k-\Delta k}(\Phi)$ are fitted 
by the polynomial \eq{polpot} in the interval $\Phi\in [-\Phi_c(k),\Phi_c(k)]$ in order to read off the new couplings $v_n(k-\Delta k)$. In such a manner the behaviour of the RG trajectories can be investigated in the deep IR region. This numerical procedure generally converges
for sufficiently small values of the ratio $\Delta k/k$.
It was shown in Ref.  \cite{Ale1999} that for $Z=1$
 the amplitude $\rho_k(\Phi) $ of the spinodal instability is a linear function of the  homogeneous background $\Phi$,
 $2\rho_k(\Phi) = -\Phi+ \Phi_c(k)$. 
 Outside of the interval $-\Phi_c(k)\le \Phi \le \Phi_c(k)$ the dimensionful blocked potential $U_{k-\Delta k}(\Phi) $  can be taken identical to $U_{k_c}(\Phi)$. 
In the IR limit $k\to 0$ and in the interval  $-\Phi_c(0)\le \Phi \le \Phi_c(0)$ the  tree-level blocking results in the downsided parabola $\t{U}_{k\to 0} (\t{\Phi})=-\hf \t{\Phi}^2$ for the dimensionless blocked potential corresponding to  $\t{v}_1(0)=-1$, $\t{v}_n(0)=0$ for $n\ge 2$. Therefore, the dimensionful potential flatens out taking a constant value in the interval $-\Phi_c(k)\le \Phi \le \Phi_c(k)$ that represents the so-called Maxwell cut.

\section{WH RG equations for scalar  $\phi^4$ models with $O(2)$ and $U(1)$ symmetry}

\subsection{Case of $O(2)$ symmetry}\label{app:o2whrg}

Here we derive the WH RG equation for the real, two-component scalar field 
theory using the ansatz for the blocked action \eq{realaction}  exhibiting
 $O(2)$ symmetry.
 The blocking relation
\bea\label{blockO2}
e^{-S_{k-\Delta k}[\underline{\phi}]}&=&\int \mathcal{D}\phi' e^{-S_k[\underline{\phi}+\underline{\phi}']}
\eea
is the straightforward generalization of the relation \eq{block} for the 2-component scalar field. Since the WH equation is applicable only in the  LPA, the lowest order of the gradient expansion, it is sufficient to Taylor-expand the action
$S_k[\underline{\phi}+\underline{\phi}']$ in the exponent of the integrand
around the homogeneous field configuration
 $\underline{\phi}(x)=\underline{\Phi}$,
\bea
  S_k[\underline{\Phi}+\underline{\phi}']&=&S_k[\underline{\Phi}] + \hf
\int d^d x  {\underline{\phi}'}^T 
 \u{\u{  S_k^{(2)}}}[\underline{\Phi}]  \underline{\phi}'
+ \ord{ (\phi')^3},\nn
\eea
where the matrix of the second functional derivative of the blocked action
has been introduced as
\bea 
\u{\u{S_k^{(2)}}}[\underline{\Phi}] &=& \begin{pmatrix}\frac{\delta^2 S[\underline{\Phi}+\underline{\phi}']}{\delta\phi_1'(x)\delta{\phi'}_1(y)} & \frac{\delta^2 S[\underline{\Phi}+\underline{\phi}']}{\delta{\phi'}_1(x)\delta{\phi'}_2(y)} \cr
\frac{\delta^2 S[\underline{\Phi}+\underline{\phi}']}{\delta{\phi'}_2(x)\delta{\phi'}_1(y)} & \frac{\delta^2 S[\underline{\Phi}+\underline{\phi}']}{\delta{\phi'}_2(x)\delta{\phi'}_2(y)} \end{pmatrix}\biggr|_{\underline{\phi'}=0}\nn
&=&  \begin{pmatrix}  S_{11} & S_{12}\cr
S_{21}& S_{22} \end{pmatrix}\delta (x-y).
\eea
Abandoning the terms of order $\ord{\phi^{\prime 3}}$ and higher, we can 
perform the Gaussian path integral  and reduce Eq. \eq{blockO2}
to the blocking relation for the blocked action
\bea
S_{k-\Delta k}[\underline{\Phi}]& =& S_k[\underline{\Phi}]+ \frac{\hbar}{2}\tr\ln \u{\u{ S_k^{(2)}}}[\underline{\Phi}].
\eea
As it is well-known, in the limit   $\Delta k /k\to 0$ 
 the neglected  terms of higher order in $\phi'$ give vanishing contributions
and one arrives at the exact WH RG equation
\bea\label{blockO22}
  \partial_k S_k[\underline{\Phi}] &=& -\lim_{\Delta k\to 0} \frac{\hbar}{2\Delta k}\tr\ln \u{\u{  S_k^{(2)}}}[\underline{\Phi}].
\eea

In order to cast Eq. \eq{blockO22} into a more explicit form, we have to evaluate the trace log in its right-hand side. Fortunately, the matrix $\u{\u{ S_k^{(2)}}}[\underline{\Phi}]$ is diagonal in momentum space consisting of $2\times 2$
block matrices in the internal space.  For the purpose of the determination
of the elements of those block matrices for given momentum $p$
 let us make the LPA ansatz 
\bea\label{ansO2} 
S[\underline{\phi}] &=& \frac{1}{2}\sum_{a=1}^2\int \frac{d^d p}{(2\pi)^d} \phi_{a,-p}\Omega(p^2) \phi_{a,p}
 + \int d^d x U_k(\u{\phi}^T\u{\phi}  )\nn
\eea
 for the blocked action. According to this, the matrix elements are
\bea 
S_{11} &=&\Omega(p^2) + U_k'(r)+\Phi_1^2 U_k''(r), \nn
S_{22} &=&\Omega(p^2) + (U_k'(r)+\Phi_2^2 U_k''(r)),   \nn
S_{12} &=& \Phi_1\Phi_2 U_k''(r)=S_{21}, 
\eea
with $U_k'(r)=\partial_r U_k(r)$ and $U_k''(r)=\partial_r^2 U_k(r)$ and $r=\hf \u{\Phi}^T\u{\Phi}=\hf (\Phi_1^2+\Phi_2^2)$. The eigenvalues  $s_+$ and $s_-$ of the block matrices of  $\u{\u{ S_k^{(2)}}}[\underline{\Phi}]$  can be determined from the vanishing of the determinant of the corresponding eigenvalue equations $s^2 - s(S_{11}+S_{22})+ S_{11}S_{22}-S_{12}^2=0$ and are
\bea 
s_+(p) &=& \Omega(p^2)+U_k'(r)+2r U_k''(r),\nn
s_- (p)&=& \Omega(p^2) + U_k'(r).
\eea
The trace log of the matrix  $\u{\u{ S_k^{(2)}}}[\underline{\Phi}]$  is the sum of the logarithms of the eigenvalues of the matrix. The trace operation in the right-hand side of Eq. \eq{blockO22} can be carried out by summation over the internal space degrees of freedom and integrating over the modes in the infinitesimally thin momentum shell $|p|\in [k-\Delta k, k]$. Thus the  WH RG equation
\bea 
k\partial_kU_k(r)& =& -\alpha k^d\ln [s_+(k)s_-(k)]
\eea
is obtained.
 From this one obtains the WH RG equation \eq{WHO2}.

\subsection{Case of  $U(1)$ symmetry}\label{app:u1whrg}

Let us now derive the WH RG equation for the $U(1)$ symmetric model given by the ansatz \eq{skomplex}.
 Splitting the field variable at scale $k$ again into the sum of the contribution $\phi$ of the IR modes with momenta $p$ such that $|p|\le k-\Delta k$ and that of $\phi'$ of the UV modes with momenta from the infinitesimal momentum shell $k-\Delta k \le |p|\le k$, we write the blocking relation as
\bea\label{blocku1}
e^{-S_{k-\Delta k}[\phi^*,\phi]}&=&\int \mathcal{D}\phi' e^{-S_k[\phi^*+{\phi'}^*,\phi+\phi']}.
\eea
Let us  expand the exponent in the integrand of the path integral around the constant background configuration $\Phi$ as
\bea\label{expu1}
\lefteqn{
  S_k[\Phi^*+{\phi'}^*,\Phi+\phi']=S_k[\Phi^*,\Phi] }\nn
&&+ \hf
 \begin{pmatrix} {\phi'}^*,\phi'\end{pmatrix}
 \cdot {\u{\u{  S_k^{(2)} }}}[\Phi^*,\Phi] \cdot \begin{pmatrix}{\phi'}^*\cr\phi'\end{pmatrix}
+ \ord{ (\phi')^3}
\eea
with the second functional derivative matrix
\bea
  {\u{\u{ S_k^{(2)}}}}[\Phi^*,\Phi]&=&  
\begin{pmatrix}
  \fdd{S}{ {\phi'}^*_p}{ {\phi'}^*_{p'} } & \fdd{S}{ {\phi'}^*_p}{ \phi'_{p'}}\cr
  \fdd{S}{\phi'_p}{ {\phi'}^*_{p'} } & \fdd{S}{ \phi'_p}{\phi'_{p'}} 
\end{pmatrix} \nn
&=&
 \begin{pmatrix}
 S_{11} & S_{12}\cr
S_{21}&S_{22}\end{pmatrix}(2\pi)^d\delta^{(d)}(p-p'),
\eea
where
\bea
 && S_{11} = U_k''(r){\Phi^*}^2,~~
S_{22}= U_k''(r)\Phi^2,\nn
&& S_{12}=S_{21}=
 \Omega(p^2) + rU_k''(r) +U_k'(r)
\eea
and $r=\Phi^*\Phi$, the repeated prime over  $U_k$ denotes repeated derivations
 with respect to the variable $r$.

Assuming that the path integral in the right-hand side of Eq. \eq{blocku1} exhibits the trivial saddle point $\phi'={\phi'}^*=0$, the first order term  vanishes in the expansion \eq{expu1}. Neglecting the terms of  the orders higher than quadratic and  performing the Gaussian path integral, we get from \eq{blocku1} the
equation
\bea\label{blowhrelu1} \!\!\!\!
S_{k-\Delta k}[\Phi^*,\Phi] &=& S_k[\Phi^*,\Phi]+ \hf\tr\ln {\u{\u{ S_k^{(2)}}}}[\Phi^*,\Phi]
\eea
for the blocked action. Here the trace in the right-hand side is taken over
momenta from the infinitesimal momentum shell $k-\Delta k \le |p|\le k$ as well as over the internal-space matrix.  The former is trivial since $ {\u{\u{ S_k^{(2)}}}}$
is diagonal in the momentum space, so that we only need the  matrix elements $S_{11}$, etc. at momentum $p=k$.  In order to take the trace over the internal space, we perform diagonalization.
The corresponding eigenvalues $s_\pm$ of the matrix $S_{ij}$ $(i,j=1,2)$ are given by the roots of
  the second order algebraic equation $s^2 - s(S_{11}+S_{22})+ S_{11}S_{22}-S_{12}^2=0$,
\bea 
s_\pm  &=&  \frac{1}{2}\biggl\{
U_k''(r)({\Phi^*}^2+\Phi^2)
\pm\biggl\lbrack
{U_k''}^2(\Phi^2+{{\Phi^*}^2})^2 \nn
&&
+ 4\Omega^2(k^2) + 8\Omega (k^2)(U'+ r U'') \nn
&&+
 4{U'}^2 + 8r U'U'' \biggr\rbrack^{1/2}  \biggr\}.
\eea
Using the above eigenvalues, making the momentum integral over the
 infinitesimal momentum shell explicit and inserting the ansatz \eq{skomplex} we can rewrite the limit $\Delta k\to 0$ of the blocking relation \eq{blowhrelu1} as
\bea 
k\partial kU_k(r) &=& -\alpha k^d  \ln ( s_+ s_-),
\eea
where we find with trivial but somewhat lengthy algebraic manipulations that
\bea 
s_+ s_-& =&  \lbrack \Omega(k^2)+U_k'(r)\rbrack\nn
&&\times\lbrack \Omega(k^2)+U_k'(r)+2r U_k''(r)\rbrack .
\eea
Then we recover just the same  WH RG equation \eq{WHO2} for the local potential
which has been obtained for the $O(2)$ symmetric model.

\end{document}